\documentclass[12pt, draftclsnofoot, onecolumn]{IEEEtran}

\usepackage{amsfonts}
\usepackage{amsmath}

\usepackage{cite}      

\usepackage{graphicx}  

\usepackage{stfloats}  
\newcommand{\ggcmt}[1]{}
\newcommand{\comment}[1]{}
\newcommand{\mycomment}[1]{}
\newcommand{\obmcmntout}[1]{} %
\newcommand{\obmreuse}[1]{} %
\newcommand{\obmabrdgol}[1]{} %
\newcommand{\obmthg}[1]{} 
\newcommand{\omgc}[1]{} %
\newcommand{\obmre}[1]{} %
\newcommand{\obmcanbt}[1]{}  
\newcommand{\obmcn}[1]{} %

\newcommand{\figcomment}[1]{}

\begin{document}

\title{Maximizing Spectrum Availability and Exploitation: How to Maximize Spectrum Sharing Benefits to the Incumbents?}

\author{Nilesh~Khambekar,~\IEEEmembership{Member,~IEEE,}
        Chad~M.~Spooner,~\IEEEmembership{Senior Member,~IEEE,}
        and~Vipin~Chaudhary,~\IEEEmembership{Member,~IEEE}
\thanks{Nilesh Khambekar and Vipin Chaudhary are with the Department of Computer Science and Engineering, University at Buffalo, SUNY, Buffalo, New York 14260-2000}
\thanks{Chad M. Spooner is with NorthWest Research Associates, Monterey, CA.}}


\maketitle

\begin{abstract}
A significant portion of the radio frequency spectrum remains underutilized due to exclusive and static allocation of spectrum. Provisioning secondary access to the underutilized spectrum could be beneficial to the incumbents if they could gain significant value out of the fallow spectrum while ensuring protection of their primary services. From an incumbent perspective, the spectrum sharing approach needs to be non-harmful as well as efficient. In order to make spectrum sharing efficient, it is necessary to maximize the spectrum \textit{available} for secondary access as well as maximize its \textit{exploitation}. We examine the impact of the conservative assumptions that lead to lesser availability of spectrum for secondary access. The problem of joint scheduling and spectrum-access footprint allocation is at the heart of maximizing the exploitation. This problem is NP-hard and we present a suboptimal approach based on the minimal spectrum consumption cost of a spectrum-access request. In order to improve the spectrum sharing potential, we investigate the impact of the various design choices for a spectrum access mechanism (SAM). The experiments demonstrate the significance of the active role of the incumbents, the benefits of fine granular spectrum access, and the need for transceiver standards for accomplishing efficient usage of the spectrum.
\end{abstract}


\section{Introduction}

Dynamic Spectrum Access (DSA) \cite{dsa_survey} aims to share the scarce radio frequency (RF) spectrum among multiple heterogeneous networks. With exclusive and static allocation of spectrum, it has been observed that the RF spectrum is significantly underutilized in the space, time, and frequency dimensions \cite{fcc_sptf}. One of the popular approaches to DSA is Opportunistic Spectrum Access (OSA) wherein the underutilized spectrum is exploited on a secondary basis when the owners of the spectrum are not actively using the spectrum. In this regard, Federal Communications Commission (FCC) ruling allowed access to television (TV) band white spaces in November 2008 \cite{fcc_2008} and the IEEE 802.22 WRAN standard was defined in 2011 to facilitate sharing of the underutilized spectrum in the TV bands.

There are several obstacles to adoption of OSA. The incumbents are concerned about disruption or degradation of their service due to harmful interference from secondary users. From the secondary-user perspective, the availability of the secondary spectrum, throughput, and quality of service cannot be ensured. These issues affect the business prospects and discourage secondary access to the underutilized spectrum. 

To address these concerns, we need to investigate the potential of the spectrum sharing enabled by DSA and defining a SAM that can maximize exploitation of the available spectrum while providing a mechanism to ensure protection of the incumbents' uses of spectrum. We acknowledge that in order to improve spectrum sharing potential, firstly the \textit{absolute} available spectrum needs to be improved. Secondly, the \textit{recovery} of the available spectrum needs to be improved; and lastly, the \textit{exploitation} of the available spectrum needs to be improved \cite{oms_db}. 

In this paper, we employ \textit{quantified dynamic spectrum access} (QDSA) paradigm that enables defining and enforcing quantified spectrum-access footprints for the cochannel transceivers. By controlling the spectrum consumption of each of the individual transmitters and receivers, efficient spectrum sharing can be accomplished. To maximize exploitation of the available spectrum, we address the problem of joint scheduling and power allocation. The problem is NP-hard and we define a suboptimal strategy based on minimizing the consumption of spectrum while maximizing the number of scheduled spectrum-access requests and minimizing the number of harmfully interfered receivers. We perform an experimental study to investigate how spectrum availability and exploitation could be maximized by choosing the transceiver and network-design parameters. We emphasize seeking coexistence between cochannel networks, fine granular spectrum access, and enforcing transceiver-performance standards. We encourage the incumbents to play an active role and extract more value out of their spectrum investments. 
 

\subsection{Related Work}
Acknowledging the concept of receiver spectrum consumption \cite{itumetrics, gastpar}, we observed the need to separately quantify the spectrum consumed by individual transmitters and receivers when multiple heterogeneous wireless networks are sharing the spectrum in the time, space, and frequency dimensions. By discretizing the spectrum dimensions, we defined a methodology to quantify spectrum consumption spaces \cite{oms1_scq, oms1_sl}. 

In order to improve spectrum sharing, it is important to understand the weaknesses of a SAM and quantify their impact on effectively exploit the underutilized spectrum. The poor exploitation of the spectrum due to conservative spectrum-access constraints enforced by the FCC white space ruling is illustrated in \cite{berk_wsc}. In \cite{oms2_sca}, we have quantified the loss of the available spectrum due to the lack of knowledge of the RF-environment. 

To maximize the recovery of the available spectrum, it is necessary to acquire the RF environment information. In \cite{sgn_icnc}, we described algorithms to estimate the RF environment information by exploiting signal cyclostationarity. Similar to `Sensing as a Service' \cite{saas_weiss}, we separate the sensing function from the secondary user radio and apply an external RF-sensor network based infrastructure for estimating spectrum consumption in real time \cite{oms4_sce}.

Stine has proposed a model based spectrum management approach that suggests using RF-system models to capture the spectrum consumption by its transceivers and assess the compatibility between multiple uses of spectrum \cite{mbsm_stine}. The quantified spectrum access approach we apply here is based on the quantification of absolute spectrum consumption spaces \cite{oms1_sl}.  

The problem of joint scheduling and power allocation is NP-hard \cite{jcpa_nphard} and a suboptimal approach is required to maximize the exploitation of spectrum. While several approaches use the concept of dynamic interference graph, for example, \cite{intf_graph}; our approach is based on prioritizing the spectrum-access requests based on their minimal spectrum consumption costs.

In \cite{pm_wwdsa}, Marshall and Kolodzy emphasized that DSA is beneficial to the secondary users \textit{as well as} to the incumbents. In order to maximize exploitation of the underutilized spectrum, we study how regulatory parameters could be defined by the incumbents in order to maximize a SAM's ability to provision secondary access to their spectrum and protect the primary receivers from harmful interference \cite{oms3_cf1}. In comparison to this work, \textit{in the present extended paper}, we first address the problems of maximizing the available spectrum and maximizing the exploitation of the available spectrum. We investigate parameters influencing spectrum availability. For maximizing exploitation of the available spectrum, we provide a detailed mathematical treatment and illustration of the concepts applied in the proposed suboptimal algorithm.

The paper is organized as follows. In Section 2, we present the spectrum consumption quantification approach that facilitates quantified sharing of the spectrum. It also enables us to quantify the available and the exploited spectrum. In Section 3, we study how the absolute available spectrum could be maximized and investigate the impact of spectrum access parameters and RF-environment knowledge. In Section 4, we describe the QDSA paradigm for defining and enforcing quantified spectrum-access footprints. In order to maximize the exploitation of the available spectrum, we present a sub-optimal approach to scheduling spectrum-access requests based on the spectrum consumed by the transceivers of the candidate spectrum-access requests. In Section 5, we perform an experimental study and investigate how the spectrum availability and exploitation could be maximized by choosing certain spectrum-access and transceiver design parameters. In particular, we develop a baseline experiment and illustrate incremental performance improvement with modifications to the baseline experiment. Finally, we draw conclusions in Section 6.

\section{Spectrum Consumption Quantification Methodology}

\subsection{How Spectrum is Consumed?}
Traditionally, we assume that spectrum is consumed by transmitters; however, the spectrum is \textit{also} consumed by receivers. Receivers consume spectrum in terms of constraining the spectrum-access by other transmitters. For example, constraints are imposed in terms of guard-bands, separation distances, and operational hours for protecting receivers. Thus, the presence of receivers enforces limits on the interference-power in the space, time, and frequency dimensions. Traditionally, the spectrum allocation is static and exclusive; hence, receivers' consumption of spectrum need not be separately considered \cite{itu}.

Dynamic spectrum sharing approach is a paradigm shift from the conventional static and exclusive approach to spectrum allocation. The networks could be \textit{spatially overlapping} and the spectrum access is \textit{shared} in the time, space, and frequency dimensions. We need the ability to quantify the spectrum that is \textit{actually} used by the transceivers of all the networks and the harmful interference caused by and to each of the transceivers\footnote{Some of the recently proposed metrics \cite{fccmetrics, xgmetrics, tandra_metrics} that can be applied have applications in specific cases and cannot be used for quantifying the utilization, availability, and degradation of spectrum under generic spectrum sharing scenarios.}. In this regard, we defined a methodology to quantify the spectrum consumption spaces by discretizing the spectrum access dimensions \cite{oms1_sl}. Here, in the context of defining a suboptimal approach to the joint scheduling and spectrum-footprint allocation problem, we provide a brief overview of the quantification methodology.

\subsection{Spectrum Consumption Quantification Methodology}

\noindent
\subsubsection{System Model}
We consider a generic collection of multiple heterogeneous RF-systems in two dimensional geometric space. Here, we define a \textit{network}\footnote{In order to make an efficient use of spectrum, we seek to define spectrum-access policy at the lowest granularity of spectrum-use within a RF-system.} as a set of a transmitter and its receivers corresponding to a spectrum-access request. 

\noindent
\textbf{Interference Model:}
\noindent
The power received from a transmitter $t_n$ of network $n$ at a point $\rho$ in the geometric space is given by 
\begin{equation}
\label{eq:gen_rcvd_power}
P(t_n, \rho) = P_{t_n} min\Big\{1, L({d(t_n,\rho)}^{\alpha})\Big\}
\end{equation}
where $P_{t_n}$ is the transmit power of the transmitter and $d(t_n,\rho)$ is the distance between the transmitter $t_n$ and the point $\rho$ in the space. ${\alpha}$ is the path-loss exponent and it is assumed that $\alpha > 2$.  $L({d(t_n,\rho)}^{\alpha})$ denotes the path-loss factor. The \textit{min} operation in ~(\ref{eq:gen_rcvd_power}) ensures that the received power is never more than the transmitted power.

We assume that transceivers can optionally employ directional transmission and reception in order to minimize interference. A receiver can withstand non-zero interference when the received Signal to Interference and Noise Ratio (SINR) is greater than the given threshold\footnote{The threshold, $\beta_{n,m}$, represents the quality of the $m^{th}$ receiver of the $n^{th}$ network and incorporates receiver-noise and other receiver technology imperfections.} $\beta_{n,m}$. 

\textit{We divide the spectrum space into multiple unit spectrum-spaces and quantify the spectrum consumption by each of the transceivers at a sample point in each of the unit spectrum-spaces.} We identify multiple spectrum consumption spaces: spectrum space consumed by a transmitter, spectrum space consumed by a receiver, spectrum space consumed by a spectrum-access request, and the spectrum space available for consumption. The spectrum consumption space in a geographical region is quantified by summing up the spectrum consumed in the unit spectrum-spaces.

We define the maximum power at any point in the system to be $P_{MAX}$ and the minimum power be $P_{MIN}$. In practice, transmitter design factors and human safety conditions determine $P_{MAX}$, and $P_{MIN}$ is driven by minimum measurable power. $P_{MAX}$ and $P_{MIN}$ together enable us to quantify the spectrum consumed by a transmitter or receiver in \textit{absolute} terms in the unit spectrum space. We consider a geographical region with $\hat{A}$ unit-regions, $\hat{B}$ unit-frequency-bands, and $\hat{T}$ unit-time-quanta. Thus, the \textit{total spectrum space} is $(P_{MAX}-P_{MIN})\hat{A}\hat{B}\hat{T}$ \ $Wm^2$.

\subsubsection{Quantifying Spectrum Consumed by a Transmitter}
We define \textit{transmitter-occupancy} in a unit-region $\chi$, in the time-quantum ${\tau}$, and in the frequency-band $\nu$ as the aggregate power received \textit{at a sample point} $\rho_0 \in \chi$ in the unit-region. Therefore, from (1), we get 
\begin{equation}
\zeta(t_n, \chi, \tau, \nu) = P(t_n, \rho_{0}, \tau, \nu) .
\end{equation}

The spectrum consumed by transmitter $t_n$ in a geographical region is the sum of the transmitter-occupancy in all the unit-regions across the temporal and spectral dimensions
\begin{equation}
\label{eq:agsput}
\Psi_{utilized}(t_n) = \sum_{k=1}^{\hat{B}} \sum_{j=1}^{\hat{T}} \sum_{i=1}^{\hat{A}} {\zeta}(t_n, \chi_i, \tau_j, \nu_k) .
\end{equation}

\subsubsection{Quantifying the Spectrum Consumed by a Receiver}
\noindent
Let $r_{n,m}$ be the $m^{th}$ receiver of the $n^{th}$ network. The amount of interference receiver $r_{n,m}$ can tolerate is
\begin{equation}
\breve{P}_{r_{n,m}} = \frac{P(t_n, r_{n,m})}{\beta_{n,m}} - W_{r_{n,m}}
\end{equation}
where $W_{r_{n,m}}$ is the average ambient noise power at $r_{n,m}$. We define this quantity as the \textit{interference-margin} at the receiver. The unit of interference-margin is $W$.

We can view interference-margin $\breve{P}_{r_{n,m}}$ as the upper-bound on the transmit-power of an interferer positioned at a spatial separation of zero. We quantify the limit on interference-power at a point $\rho$ in space in terms of the receiver-imposed interference-power upper bound.
\begin{equation}
\label{eq:riub}
\breve{I}(r_{n,m}, \rho) = \breve{P}_{r_{n,m}} min\{1, L(d(\rho, r_{n,m})^{\alpha})\}
\end{equation}
where $d(\rho, r_{n,m})$ is the distance between the receiver $r_{n,m}$ and the point $\rho$ in space. $L(d(\rho, r_{n,m})^{\alpha})$ denotes the distance based path-loss.

Let $\ddot{I}(r_{n,m}, \rho)$ represent the aggregate interference seen by receiver $r_{n,m}$. Then, the \textit{interference opportunity} imposed by this receiver is given by the difference between the upper bound on the interference and the aggregate interference already experienced,
\begin{equation}
\label{eq:spoppt}
\acute{I}(r_{n,m}, \rho) = \breve{I}(r_{n,m}, \rho) -  \ddot{I}(r_{n,m}, \rho)
\end{equation}

Let \textit{receiver-liability} represent the spectrum consumed by a receiver at a point in the spectrum-space. The receiver-liability of the receiver $r_{n,m}$ at point $\rho$ is the difference between the maximum power at any point  and the maximum power allowed by the receiver $r_{n,m}$ at point $\rho$. The maximum power allowed by a receiver at a point is the sum of spectrum occupancy and interference opportunity. Therefore,
\begin{equation}
\label{eq:sprlpt}
\phi(r_{n,m}, \rho) = P_{MAX} - (\omega(\rho) + \acute{I}(r_{n,m}, \rho)) ,
\end{equation}
where $\omega(\rho)$ is the \textit{spectrum-occupancy} and is defined as the aggregate power received at a point $\rho$ in space
\begin{equation}
\label{eq:spocpt}
\omega(\rho) = \sum_{n}	P(t_n, \rho) + W_{\rho} .
\end{equation}

The \textit{receiver-liability} in a unit-region $\chi$, at the ${\tau}^{th}$ snapshot of time, and in the frequency-band $\nu$,  is defined as the receiver-liability \textit{at a sample point} $\rho_0 \in \chi$ in the given unit-region 
\begin{equation}
\label{eq:spfb}
{\Phi}(r_{n,m}, \chi, \tau, \nu) = \phi(r_{n,m}, \rho_0, \tau, \nu) . 
\end{equation}

The \textit{forbidden spectrum} in a geographical area is quantified as the sum of receiver-liability in all the unit-regions across all the frequency bands of interest, in the $\hat{T}$ time-quanta. Therefore, 
\begin{equation}
\label{eq:agspfb}
\Psi_{forbidden}(r_{n,m}) = \sum_{k=1}^{\hat{B}} \sum_{m=1}^{\hat{T}} \sum_{i=1}^{\hat{A}} {\Phi}(r_{n,m}, \chi_i, \tau_j, \nu_k) .
\end{equation}

\subsubsection{The Available Spectrum Space}

By spatially combining the limits on the maximum interference power imposed by all the receivers of all the scheduled spectrum-access requests, we obtain \textit{spectrum opportunity} at a point $\rho$ as
\begin{equation}
\gamma(\rho)	= \min_n (\min_m (\acute{I}(r_{n,m}, \rho)))
\end{equation}

The spectrum-opportunity at a point represents the maximum transmit-power that can be assigned to a potential transmitter without causing harmful interference to any of the cochannel receivers.

The spectrum-opportunity in a unit-region $\chi$ at the ${\tau}^{th}$ snapshot of time, and in the frequency-band $\nu$, is defined as the spectrum-opportunity \textit{at a sample point} $\rho_0 \in \chi$ in the given unit-region. 
\begin{equation}
\label{eq:urspop}
{\Gamma}(\chi, \tau, \nu) = \gamma(\rho_0, \tau, \nu) .
\end{equation}
To quantify the available-spectrum in a geographical region, we need to sum spectrum-opportunity in all the unit-regions across the temporal and spectral dimensions
\begin{equation}
\label{eq:agspav}
\Psi_{available} = \sum_{k=1}^{\hat{B}} \sum_{j=1}^{\hat{T}} \sum_{i=1}^{\hat{A}} {\Gamma}(\chi_i, \tau_j, \nu_k) .
\end{equation}

\subsubsection{Network Spectrum Consumption}

The spectrum consumed by a network is quantified by summing up the spectrum consumption by its transmitter and the receivers.

\begin{equation}
\label{eq:nsc}
\Psi_{NSC}(n) = \Psi_{utilized}(t_{n}) + \sum_m \Psi_{forbidden}(r_{n,m}) 
\end{equation}

\textit{Minimum network spectrum consumption:}
The spectrum consumption by receivers is minimum when receivers do not receive any interference power, that is the spectrum consumption by a transmitter and its receivers is minimum when the spectrum access is not shared with any other networks. We term it \textit{minimum network spectrum consumption} and apply it in the algorithm for the joint scheduling and spectrum-footprint allocation problem in Section 4.

\subsubsection{Illustration: Network Spectrum Consumption}
\label{illn_setup}
 
Let us quantify\footnote{Due to space, we cannot illustrate the spectrum consumption quantification methodology in detail. Refer to \cite{oms1_sl} for detailed illustration and extensive treatment of spectrum-space discretization.} the spectrum consumed by an example network in 4.3 $km$ x 3.7 $km$ geographical region. Let us assume that the region is divided into hexagonal unit-region with 100 $m$ side. Let $P_{MAX}$ to be 30 dBm, an arbitrary high value and $P_{MIN}$ to be -192 dBm, an arbitrary low value. Considering spectrum sharing in a unit time-quantum and a unit bandwidth channel, the total spectrum space is $676$ $Wm^2$. 

\begin{figure}[htbp!]
\centering
{\includegraphics [width=0.472\textwidth, angle=0] {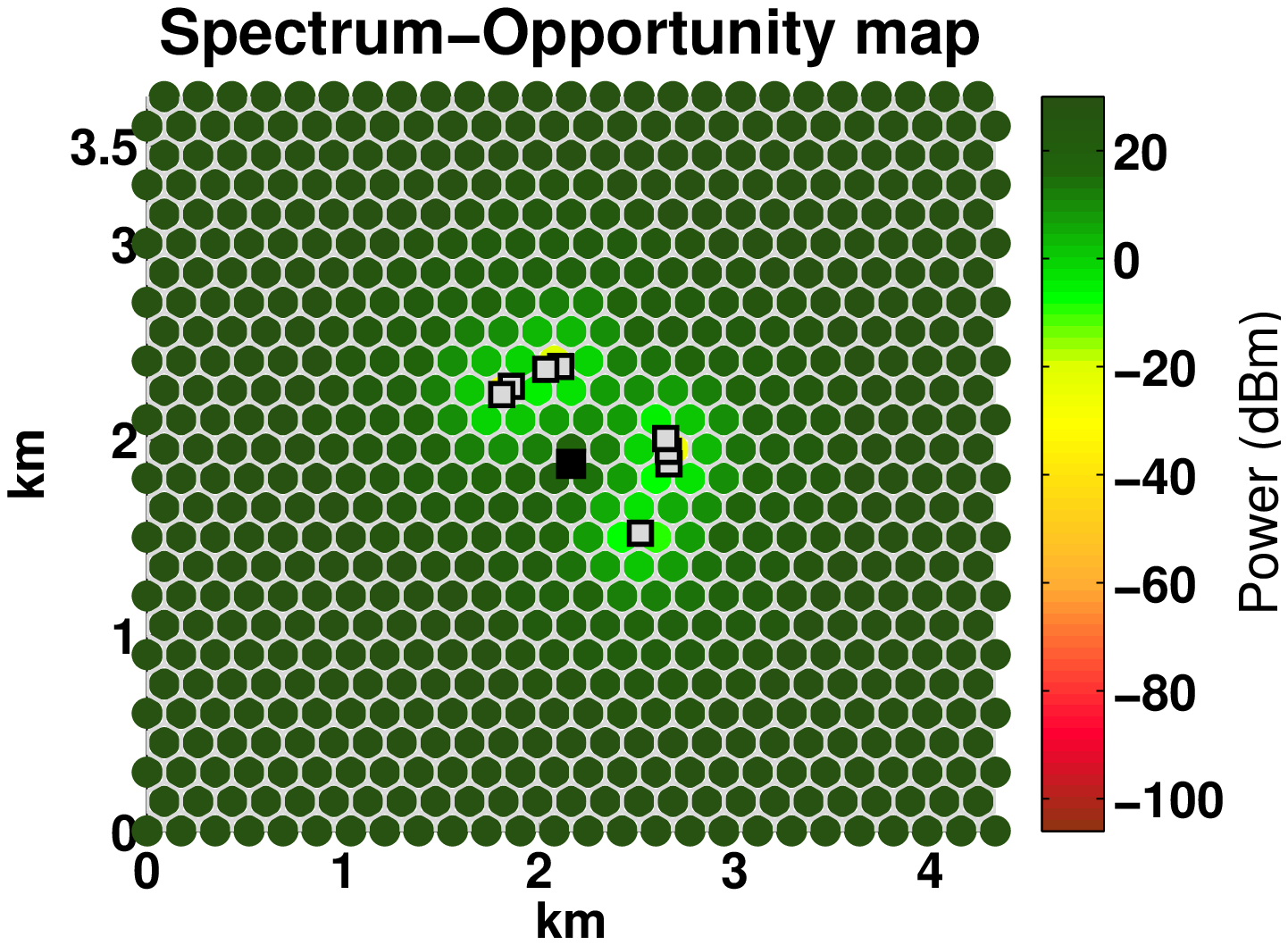}}
{\includegraphics [width=0.472\textwidth, angle=0] {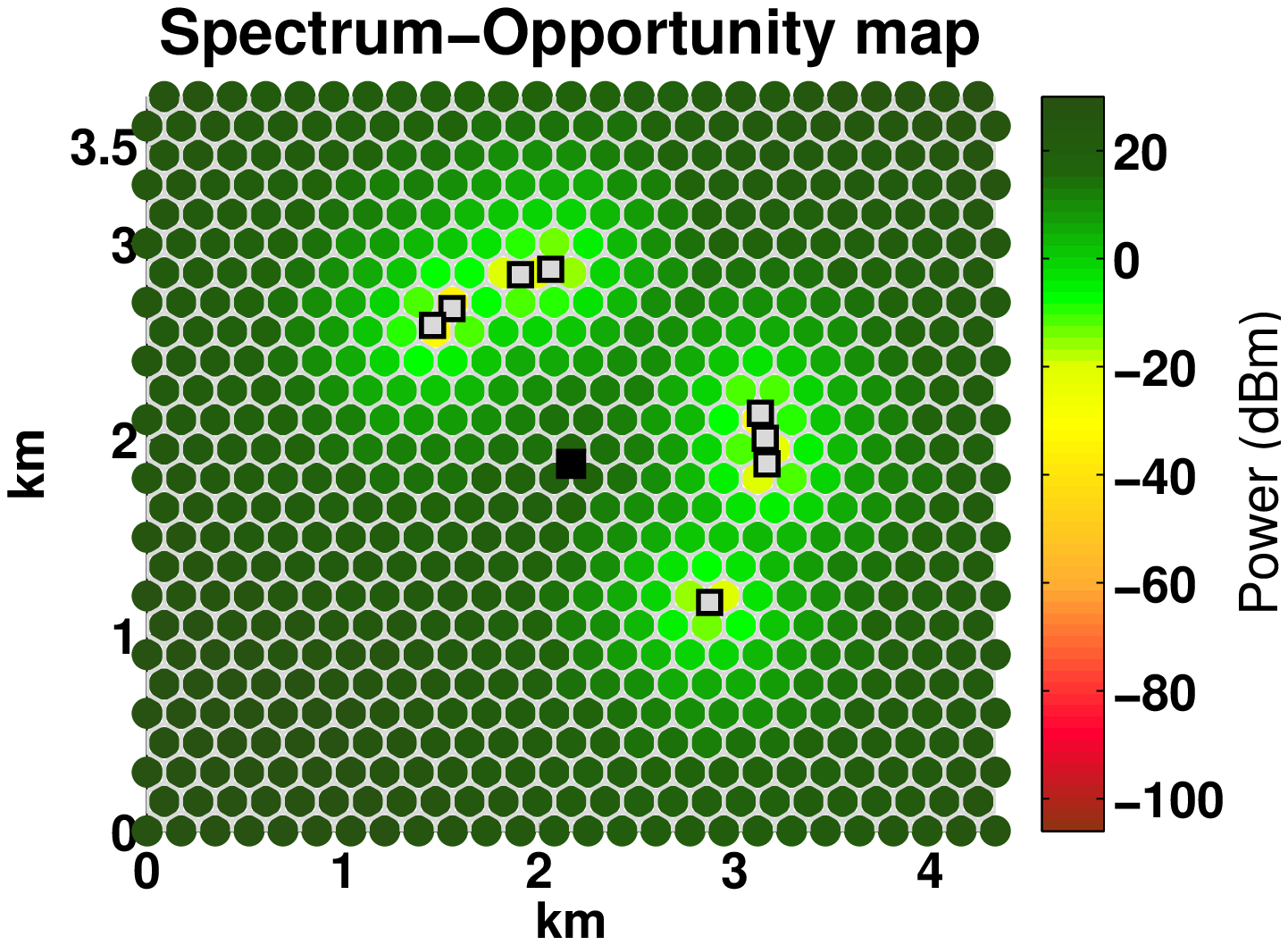}}
{\includegraphics [width=0.472\textwidth, angle=0] {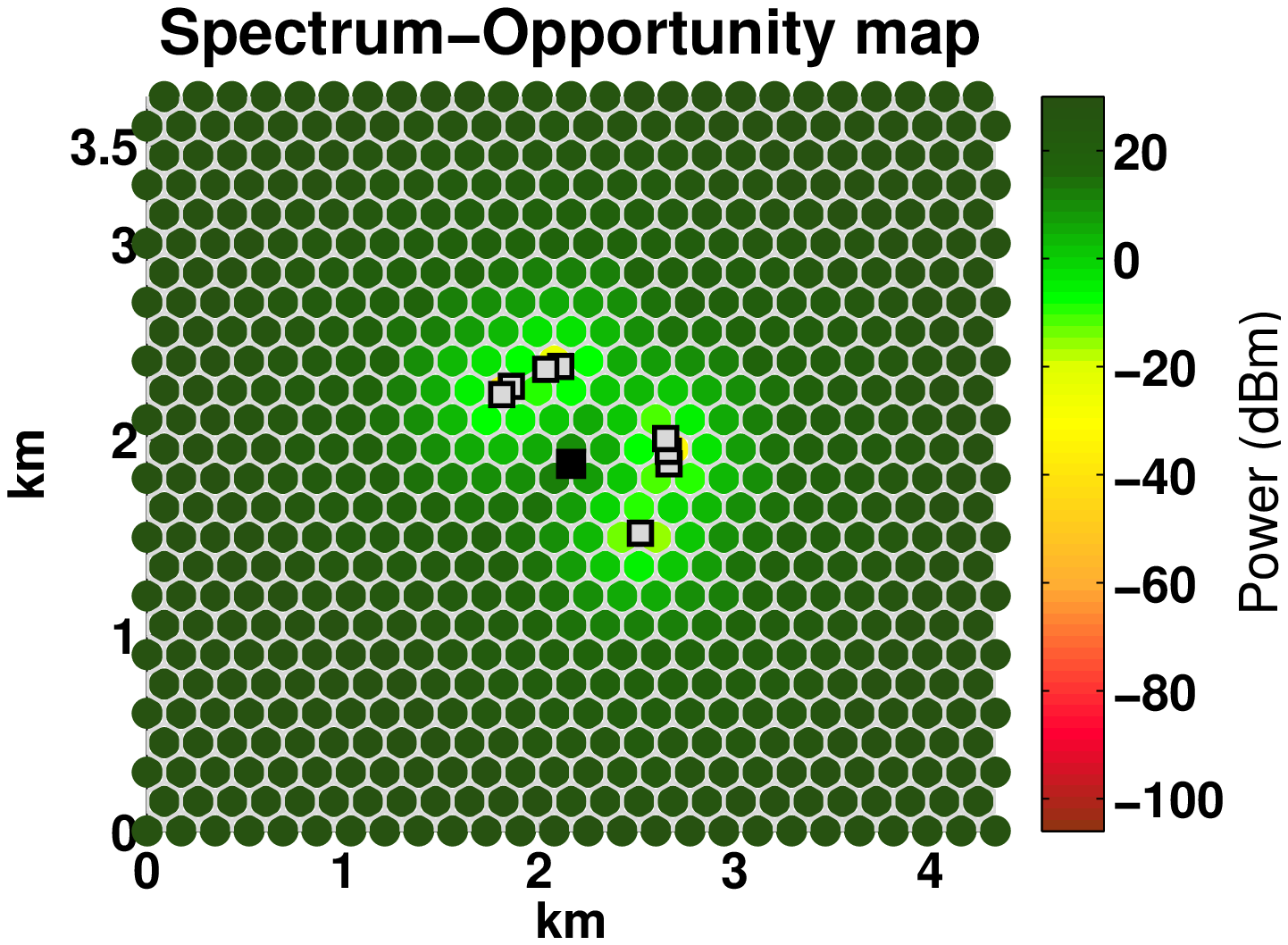}}
{\includegraphics [width=0.472\textwidth, angle=0] {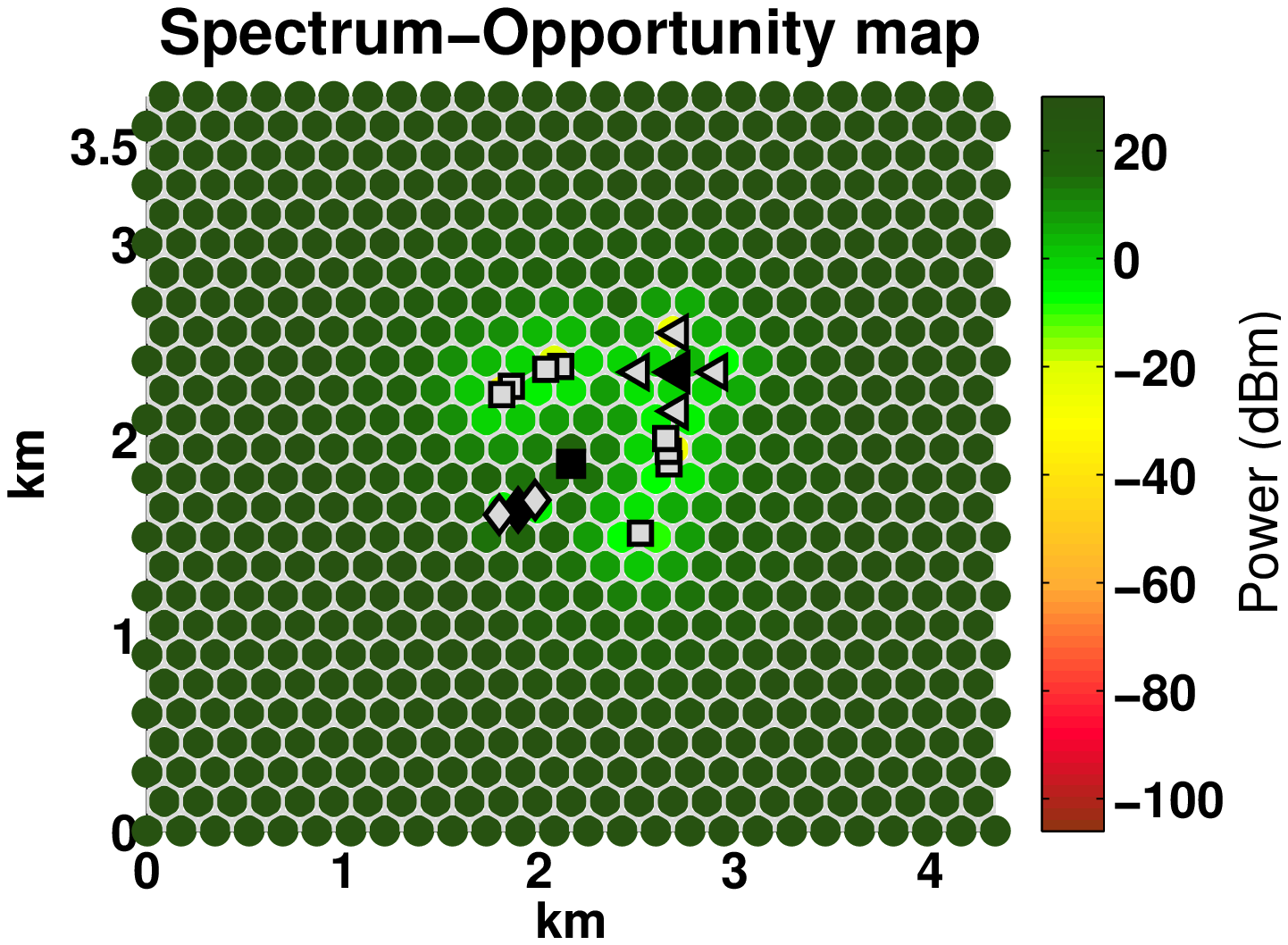}}
\caption{The figure illustrates the consumption of spectrum by a network under four different scenarios. Each plot shows spatial distribution of the spectrum-opportunity.  A transmitter is shown with a solid shape and its receivers are shown with the same non-solid shape. The higher the spectrum-opportunity in a unit-region, the lower is the spectrum consumption by the transmitters and the receivers in that unit-region. The aggregate spectrum consumption by the network with square shape in the given geographical region in the four cases is as follows: 185.3 $Wm^2$ i.e. 27.4\% (top-left), 564.1 $Wm^2$ i.e. 83.4\% (top-right), 337.6 $Wm^2$ i.e. 49.9\% (bottom-left), and 191.8 $Wm^2$ i.e. 28.9\% (bottom-right). With increased distance between the transmitter and the receivers, the network spectrum consumption is higher; with reduced transmit-power, the network spectrum-consumption is higher. Finally, the network spectrum consumption depends on the interference the network-receivers experience.}
\label{fig:L205_NSC_illn}

\end{figure}

Let us consider the topology in the \textbf{\textit{top-left}} portion of Figure~\ref{fig:L205_NSC_illn}. Spectrum-access is exercised by a transmitter and its eight receivers. The transmitter is at the center of the geographical region and all eight receivers are at a distance of 500 $m$ from the transmitter. The transceiver antennas are omnidirectional. The transmitter power is 21 dBm and the minimum SINR for successful reception is 3 dB. The propagation conditions are simplistic, with only distance-dependent path-loss. The path-loss index is assumed to be $3.5$. The noise-floor is assumed to be -106 dBm corresponding to a 6MHz channel bandwidth. In this case, the inband SINR at the receivers is 32.6 dB and the spectrum consumed by the transmitter and its receivers is 185.3 $Wm^2$ (27.4\% of the total spectrum space). In the \textbf{\textit{top-right}} topology shown in Figure~\ref{fig:L205_NSC_illn}, the receivers are at 1000 $m$ distance from the transmitter. The receivers experience a SINR of 22.1 dB and the spectrum consumption by the transceivers is 564.1 $Wm^2$ (83.4\% of the total spectrum space). In the \textbf{\textit{bottom-left}} topology of Figure~\ref{fig:L205_NSC_illn}, the transmitter is assumed to exercise 6 dB lower transmit-power of 15 dBm and the receivers are again 500 $m$ from the transmitter. The SINR at the receivers is 26.6 dB and the network spectrum consumption is 337.6 $Wm^2$ (49.9\% of the total spectrum space). We note that even though the spectrum consumed by the transmitter is reduced, the receiver spectrum consumption has in fact significantly increased due to lower SINR at the receivers. \textit{As a result, the spectrum available for sharing with other spatially overlapping networks is reduced. }

We note that in the three topologies considered so far, the receivers do not receive any interference from cochannel transmitters. Next, we assume that two additional networks are exploiting spectrum-access as shown in the \textbf{\textit{bottom-right}} topology in Figure~\ref{fig:L205_NSC_illn}. In this case, none of the receivers of the three networks experience harmful interference. (The SINR at all the receivers is higher than the minimum desired SINR for successful reception of the signal.) The spectrum consumed by all the cochannel transmitters and receivers is 195.2 $Wm^2$ (28.9\% of the total spectrum space). The spectrum consumed by the network with transmitter at the center is 191.8 $Wm^2$ (28.4\% of the total spectrum space). Here, we observe that with spectrum consumption by the network is higher as compared to the previous case (27.4\%) as the receivers experience some interference power from the cochannel networks and experience lower SINR.

\section{How to Maximize the Available Spectrum Space}
\noindent
For maximizing the spectrum sharing potential, we need to maximize the available spectrum space and the exploitation of the available spectrum space. In this section, we focus on the first objective and investigate the key parameters that influence the available spectrum space. In this regard, we quantify the impact of primary user (PU) transmit-power, lack of knowledge of the receiver positions, and the path-loss exponent (PLE) uncertainty.

\subsection{Impact of PU Transmit-power on the Available Spectrum Space}
\begin{figure}[htbp!]
\centering
{\includegraphics [width=0.49\textwidth, angle=0] {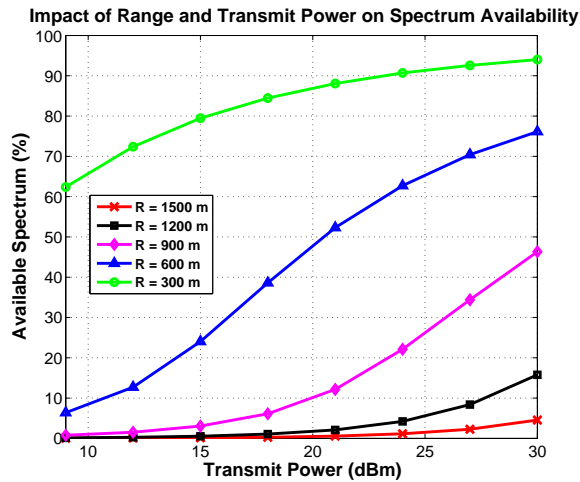}}
\caption{The plot shows the available spectrum space increases with the \textit{PU transmit-power} due to increase in the SINR at the PU receivers in accordance with (\ref{eq:urspop}) and due to reduction in the range of the incumbent RF-system. The available spectrum space increases with lower value of the range (R) of the network.}
\label{fig:L306_AV}
\end{figure}

When the transmit-power employed by the incumbent network is increased, the PU receivers experience higher SINR, and the available spectrum space is significantly increased. Figure \ref{fig:L306_AV} shows the impact of the transmit-power of a PU transmitter on the available spectrum space. It also captures the impact of the service-range of the incumbent RF-system. \textit{We argue that incumbents need to employ higher transmit-power in order to maximize the spectrum available for secondary access.}

\subsection{Impact of Lack of Knowledge of the Receiver Positions on the Available Spectrum Space}
\begin{figure}[htbp!]
\centering
{\includegraphics [width=0.49\textwidth, angle=0] {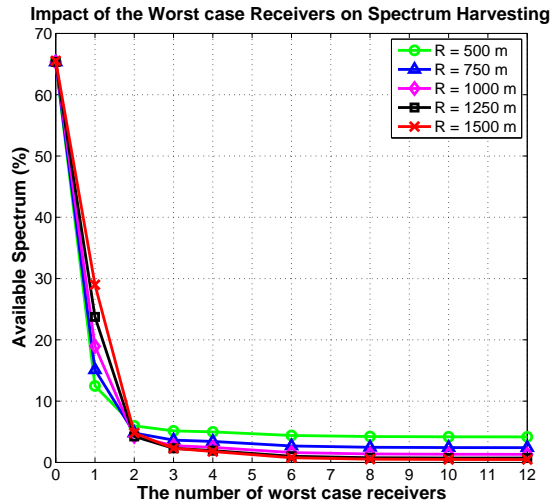}}
\caption{The figure shows that the receivers at the worst-case (boundary) positions consume a significant amount of spectrum. When we lack the knowledge of the \textit{actual} receiver positions and assume all the receivers are at the boundary (R) positions, a significant amount of available spectrum space is effectively \textit{lost}.}
\label{fig:L307_AV}
\end{figure}

Due to lack of knowledge of the receiver positions, the secondary user (SU) transceivers are required to assume the worst-case positions of the primary receivers. In the next experiment, we quantify the \textit{lost available spectrum} with the assumption that the \textit{actual} receiver positions are not known and the worst-case receivers positions are considered. In this beginning, we position a PU receiver close to the PU transmitter (at a distance of 250 $m$). Then, we place an increasing number of receivers at the periphery of the network and quantify the available spectrum. From Figure \ref{fig:L307_AV}, we observe that as the number of boundary-case receivers increases, the spectrum available for sharing with other potential uses is increasingly  lost. \textit{In order to maximize the available spectrum, it is necessary to incorporate the knowledge of the actual receiver positions.}

\subsection{Impact of Conservative PLE Assumptions on the Available Spectrum Space}
\begin{figure}[htbp!]
\centering
{\includegraphics [width=0.49\textwidth, angle=0] {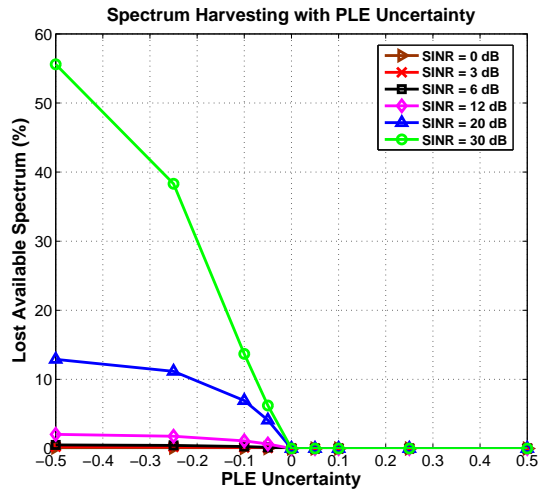}}
\caption{The plot illustrates the impact of conservative assumptions with regards to PLE. When the assumed PLE is lower than the actual PLE, we lose a significant amount of the available spectrum. The lost spectrum availability with PLE uncertainty is more pronounced when the available spectrum is higher i.e. when the receivers are experiencing higher SINR.}
\label{fig:L304_AV}
\end{figure}

In the next experiment, we quantify the \textit{lost available spectrum} due to conservative assumptions on mean PLE. From Figure~\ref{fig:L304_AV}, we observe that assuming a conservative PLE value results in a significant loss of available spectrum\footnote{On the other hand, assuming an aggressive value for mean PLE gives a false impression of higher spectrum recovery and it may lead to harmful interference at the receivers. Please refer to \cite{oms2_sca} for additional results.}. This experiment emphasizes the need for fine granular characterization of the propagation environment in order to \textit{maximize available spectrum space}. We address this problem in \cite{oms4_gl2, oms4_sce}.

\section{Maximizing the Exploitation of the Available Spectrum Space}

\mycomment{
In order to maximize the \textit{exploitation} of the available spectrum space, we need to have an effective approach for scheduling of the spectrum-access requests and defining the spectrum-access parameters. In this regard, we present a suboptimal algorithm based on the strategy of minimizing spectrum consumption. 

We apply this algorithm in the context of the quantified spectrum access paradigm that seeks the lowest (finest) spectrum-access granularity and provides the ability to \textit{precisely control} the spectrum consumption by spectrum-access requests. Here, we briefly describe the quantified spectrum access paradigm. Please refer to \cite{oms1_sl} for more details.
}
In the last section, we investigated maximizing of the spectrum available for secondary access. Having maximized the \textit{absolute} available spectrum space, its recovery needs to be maximized. We address this problem by learning RF-environment using an external RF-sensor network and inferring the spectrum consumption information in \cite{oms4_sce}. Thirdly, we need to maximize the exploitation of the \textit{recovered} available spectrum. In this regard, we present the QDSA paradigm \cite{oms1_sl} that seeks the lowest (finest) possible spectrum-access granularity and provides the ability to \textit{precisely control} the spectrum consumption by spectrum-access requests.  

\subsection{Quantified Dynamic Spectrum Access (QDSA) Paradigm}
The QDSA paradigm enables \textit{quantified} sharing of the spectrum resource among spatially overlapping multiple heterogeneous RF-systems. The consumption of the spectrum by the transceivers is quantified in terms of \textit{unit-regions} in the spatial dimension, \textit{unit spectral-bands} in the frequency dimension, and \textit{unit time-quanta} in the temporal dimension. A \textit{spectrum-access footprint} of a transmitter or a receiver represents the spectrum consumed in the space, time and frequency dimensions. A \textit{spectrum-access policy} represents the spectrum-access attributes given the transmitter and receiver positions, the propagation environment, and the expected link quality. There are following essential spectrum management functions:
\begin{enumerate}
	\item Estimating the available spectrum space,
	\item Spectrum assignment that is assigning spectrum-access footprints to individual transceivers of a spectrum-access request
	\item Defining a dynamic spectrum-access policy, and
	\item Enforcement of the spectrum-access policy.
\end{enumerate}

For accomplishing the first and the last functions, the propagation environment needs to be characterized and the spectrum consumption spaces need to be estimated in real time. This can be accomplished with a grid of RF-sensors. Having known the available spectrum, the spectrum consumed by each of the transmitters and receivers is precisely controlled in terms of the dynamic spectrum-access policy.

The QDSA paradigm provides a SAM the visibility into the absolute available spectrum space and the spectrum consumed by the individual transceivers under the interference management approach chosen by the SAM. It additionally provides the capability to regulate the spectrum access provisioned by a SAM.

\subsection{Scheduling Spectrum-access Requests and Assigning the Spectrum-Access Footprints}

\subsubsection{Problem Description}
A SAM needs to schedule a set of spectrum-access requests and assign a spectrum-access footprint to each of the scheduled requests given a pool of the available spectrum resource. It needs to choose a frequency channel, the transmit-power, the directionality of the transceivers, and the spectrum-access time interval while ensuring no harmful interference to the cochannel receivers. In its simplest form, the problem of joint scheduling and defining spectrum access footprint reduces to the problem of \textit{joint scheduling and power allocation}. This problem is known to be NP-hard \cite{jcpa_nphard}; Hence, we need to define a suboptimal approach.  

\subsubsection{Considerations in Scheduling and Defining of Spectrum-Access Footprints}
In order to define a suboptimal approach, we discuss the preferred SAM attributes and the intended behavior of the algorithm.\\
\noindent
\textbf{Preferred SAM attributes}\\
\noindent
We note that a SAM that \textit{minimizes the consumption of the spectrum} while maximizing the number of scheduled spectrum-access requests and minimizing the number of harmfully interfered receivers could be beneficial from a spectrum-provider's or an incumbent's perspective. The excess spectrum could be applied for scheduling a future set of spectrum-access requests, to handle user mobility, or to provide spectrum redundancy. 

\noindent
\textbf{Choosing Transmit-Power for a Spectrum-Access Footprint}\\
From (4) and (11), we note that the higher the transmit-power, the higher is the interference-margin at the receivers, and therefore, higher is the spectrum-opportunity across the unit-regions of the geographical region. However, if the transmitter is located close to a cochannel receiver, the receiver may get harmfully interfered. A SAM having the knowledge of the receiver positions can avoid scheduling the spectrum-access-request or can schedule the spectrum-access-request with lower transmit power. However, the latter has the effect of reducing SINR at a receiver and consequently, it lowers the spectrum-opportunity across all the unit-regions of the geographical region. In this case, as more spectrum-access requests are scheduled, the cochannel receivers may experience harmful interference. Thus, depending upon the strategy for interference management of the SAM, employing a lower secondary user transmit power may lead to one of two effects
\begin{itemize}
	\item The number of harmfully interfered SU-receivers increases
	\item The number of scheduled spectrum-access requests decreases.
\end{itemize}
On the other hand, boosting the transmit power reduces the spectrum consumption by the receivers. The interference-opportunity imposed by a receiver is higher than the transmitter-occupancy by a transmitter as the receiver quality factor, $\beta$, from (1) is greater than unity. Effectively, network spectrum consumption decreases with the increase in transmit-power. The effect is more pronounced with higher number of receivers within a network.

\subsubsection{Scheduling by Minimal Network Spectrum Consumption}
In this paper, we propose a suboptimal strategy to \textit{minimize the spectrum consumed} by the scheduled spectrum-access requests. First, we quantify the \textit{minimal} spectrum consumed by a candidate spectrum-access request given by (\ref{eq:nsc}) and next schedule spectrum-access requests in the ascending order of their \textit{minimal network spectrum consumption} costs while ensuring the minimum SINR constraint is satisfied for \textit{all} the receivers. We term this approach as Network Spectrum Consumption based Coexistence (NSC-CX). Following are the steps:
\begin{enumerate}
  \item Choose a channel for scheduling requests from the pool of the available channels.
	\item Quantify the \textit{minimal} spectrum consumed by each of the candidate requests.
  \item Schedule the request with the lowest minimal network spectrum consumption cost if it does not cause harmful interference to the receivers of the previously scheduled spectrum-access requests. If it causes harmful interference to any of the receivers, the spectrum-access request is excluded. 
  \item Repeat steps 2 and 3 until all requests are considered. 
  \item Repeat steps 1, 2, 3, and 4 for all the unscheduled requests.
\end{enumerate}

\section{Maximizing Spectrum Sharing Benefits to the Incumbents}

In this section, we investigate the impact of various SAM design choices, the capabilities of transceivers, and granularity of spectrum access with regards to maximizing the availability and exploitation of the spectrum. First, we define a base-case simulation experiment and compare performance of various SAMs. We iteratively modify the experiment and observe the potential benefits for the spectrum-owners.

\subsection{Setup}
We consider a 4.3 $km$ x 3.7 $km$ geographical region with a primary transmitter at the center. We discretize the spectrum-space as previously described in Section \ref{illn_setup} and the total spectrum is 676 space $Wm^2$. We assume SUs do not have knowledge of the positions of the PU receivers and  are assumed to be at the worst-case positions. We consider distance-dependent path loss model with the mean path-loss exponent as 3.5. The initial parameter settings are as follows.
\begin{itemize}
	\item The minimum desired SINR for the worst-case PU receivers is 20 dB and the SINR experienced is equal to the minimum desired SINR, i.e. 20 dB. The  minimum desired SINR for secondary receivers is 3 dB.
	\item The range of the PU RF-system is 500 $m$. The range of the SU networks is 100 $m$ and the SU networks are scattered in the geographical region with uniformly distributed SU transmitter positions. 
	\item All the PU and SU transceivers are employing omnidirectional antennas.  
\end{itemize}
We \textit{incrementally} change these parameters in order to reduce the spectrum consumption  by transceivers and increase the \textit{available} spectrum space in the following series of experiments.  This helps us study how the available spectrum gets \textit{exploited} by the candidate SAMs. In an actual scenario, RF-sensor network needs to be deployed for recovering the available spectrum space; in the experimentation, the available spectrum space is quantified based on the spectrum-access parameters of the transceivers and the modeled propagation environment. Regarding the size of geographical region, we purposefully choose a relatively smaller region as our emphasis is on exploiting the fine granular spatial spectrum opportunities. We deploy 100 SU networks within this geometry. A larger region would require a significantly higher number of SU networks to exploit the fine granular opportunities and the simulation complexity grows several folds. 

\subsection{The SAM Candidates}
Per the \textbf{Underlay} SAM \cite{dsa_survey}, the secondary users (SUs) exploit the spectrum with a very low transmit-power in order to not cause severe interference at the primary user (PU) receivers. We consider the secondary transmit-power to be 30 dB above the thermal noise power (-106 dBm for 6 MHz band). The Underlay SAM does not require to check whether the primary RF-system is active.

The second approach to spectrum access is the \textbf{Overlay} SAM which requires secondary user devices to confirm that the primary transmitter signal is not present before it can access the spectrum with constrained transmit-power \cite{dsa_survey}. The key concern with this approach is the sensitivity required for PU detection translates to a large spatial range \cite{oms2_sca}. Thus, in most of the spatial locations, the spectrum could not be exercised when the primary RF-system is active.  

The third approach we investigate is an enhancement to the previous overlay approach which employs a fixed limit on the SU  transmit-power. It uses \textit{dynamic} SU transmit-power in order to ensure high SINR for its receivers and protect those receivers from cochannel interference from other SU networks. However, it cannot strictly ensure non-harmful interference to the primary  users due to interference aggregation effect. We term this approach Secondary Throughput-oriented OVerlay SAM \textbf{(STOV SAM)}.

The final SAM assumes knowledge of the locations of the primary receivers and thus can correctly infer the interference margin imposed by the primary receivers. However, when multiple secondary networks are exercising secondary access with transmit power implied by the interference margin, the primary receiver may still experience harmful aggregate interference. To avoid this interference aggregation scenario, a \textit{guard margin} (of 10 dB) is used when deriving the transmit power for the secondary network.  We term this approach Secondary Throughput and Primary Protection oriented OVerlay SAM \textbf{(STPPOV SAM)}.

Table 1 \mycomment{\ref{table:SAMCandSum}} shows the summary of spectrum access strategies of the SAM candidates. The four SAMs allow the SU networks to access the spectrum as long as their minimum SINR is met. To realize fairness to all the cochannel SU networks, the SU transmit power is proportionately increased until the transmit power reaches the upper limit. In addition to this, after a few experiments, we introduce a fifth SAM, NSC-CX SAM, that emphasizes non-harmful interference to all cochannel receivers and prioritizes spectrum-access requests based on the \textit{minimal network spectrum consumption} costs.

\begin{table}[h!b!p!]
\label{table:SAMCandSum}
\caption{Summary of the SAM Properties}
\centering
\begin{tabular}{lll}
\hline
SAM & Potential Interference & Constraint on SU maximum transmit-power\\
\hline
Underlay SAM  & to PU and SU & Fixed and low\\
Overlay SAM & to PU and SU & Fixed and high\\
STOV SAM  & to PU and SU & Dynamic and high\\
STPPOV SAM  & to SU only & Dynamic and high\\
\hline
\end{tabular}
\end{table}

\subsection{Metrics}
From an incumbent's or a spectrum-provider's perspective, the preferred SAM minimizes the consumption of resource while maximizing the number of scheduled spectrum-access requests and minimizing the number of the harmfully interfered receivers. We compare the performance of SAMs in terms of
\begin{itemize}
	\item the number of scheduled spectrum-access requests,
  \item the number of harmfully interfered receivers, 
  \item the percentage of the available spectrum exploited by a SAM, and
	\item the percentage of the spectrum that remained unexploited (the available spectrum).
\end{itemize}

\subsection{Experiments}

\subsubsection{The Base-Case Experiment}
\begin{figure}[htbp!]
\centering
{\includegraphics [width=0.464\textwidth, angle=0] {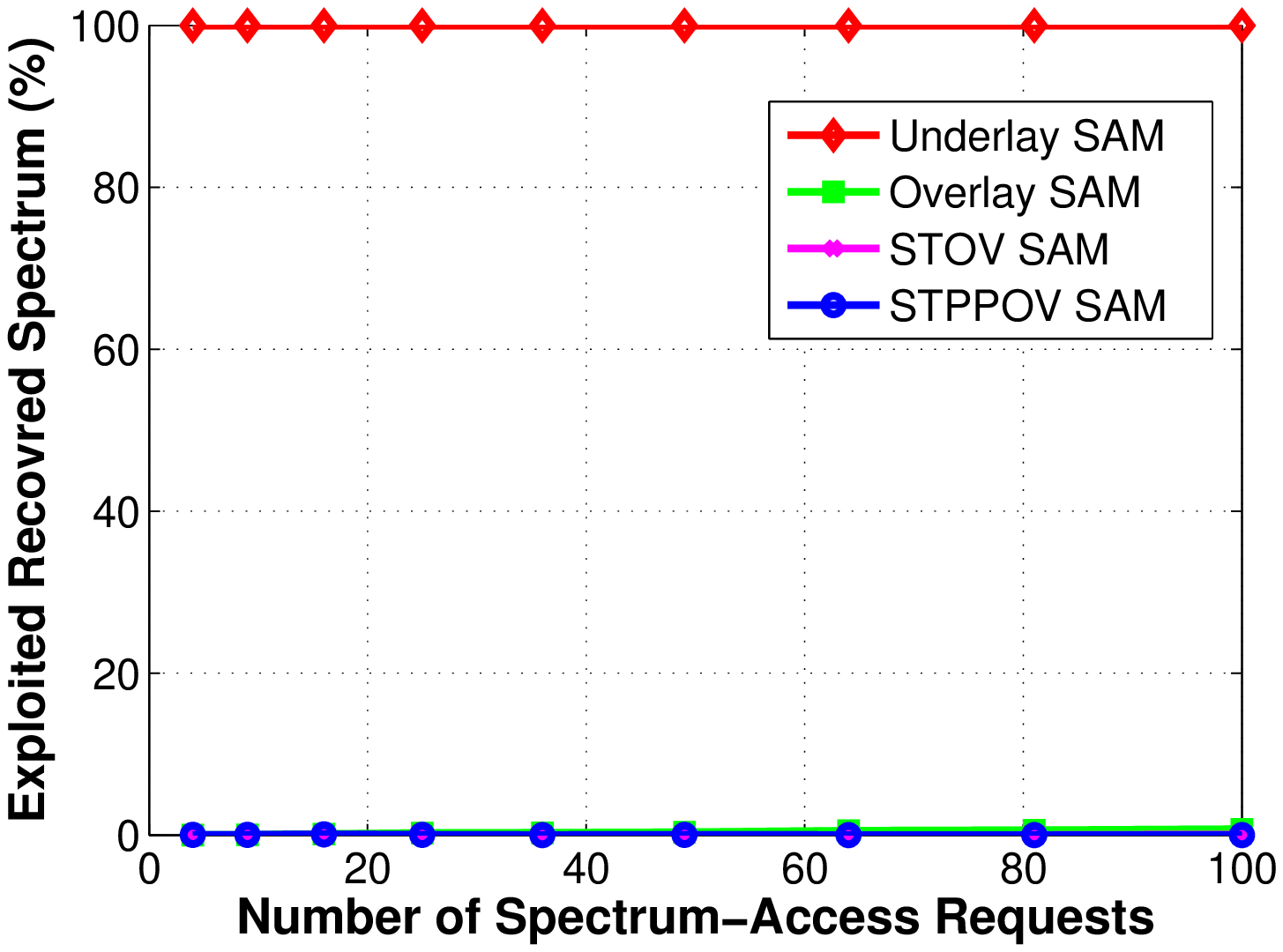}}
{\includegraphics [width=0.464\textwidth, angle=0] {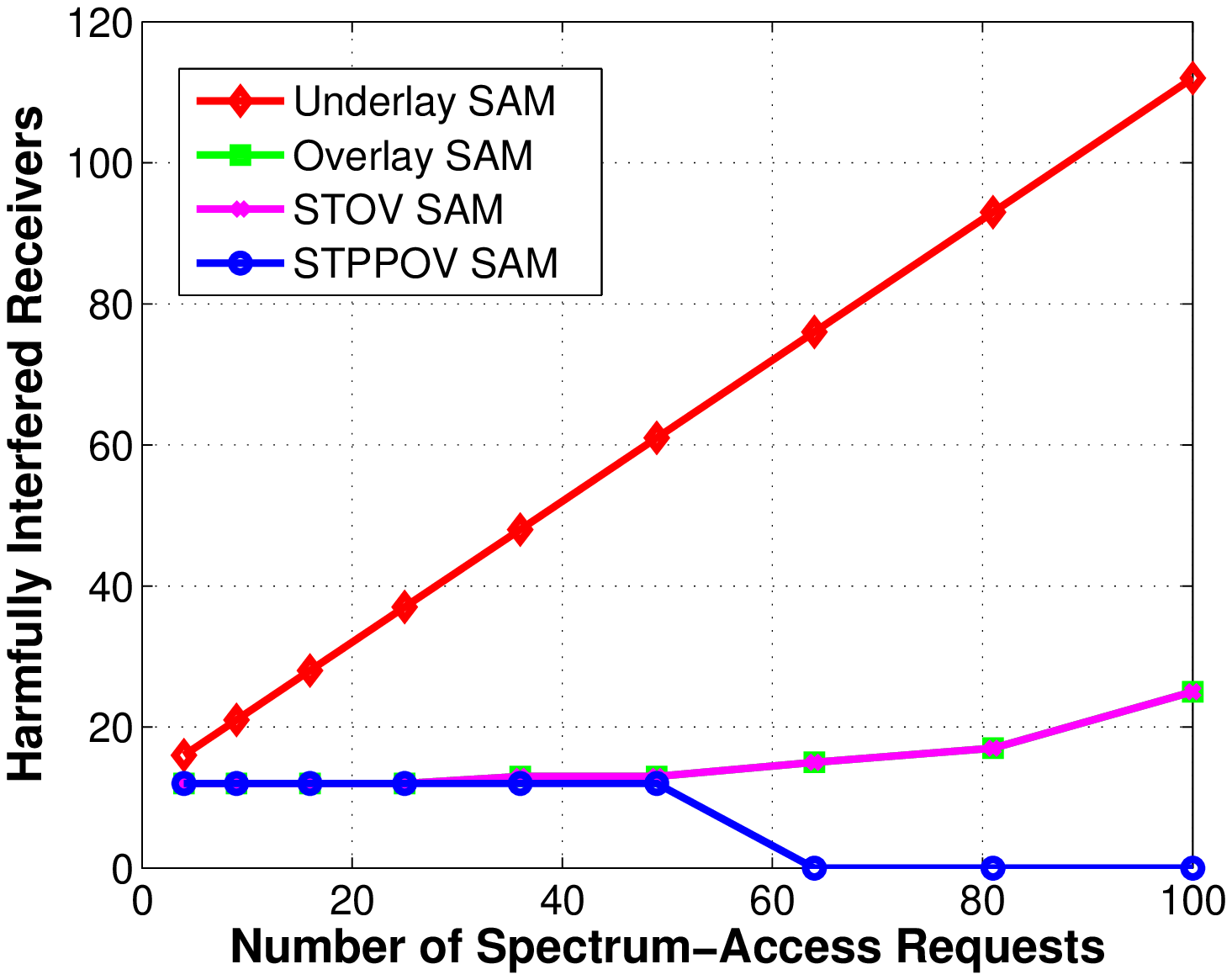}}
{\includegraphics [width=0.464\textwidth, angle=0] {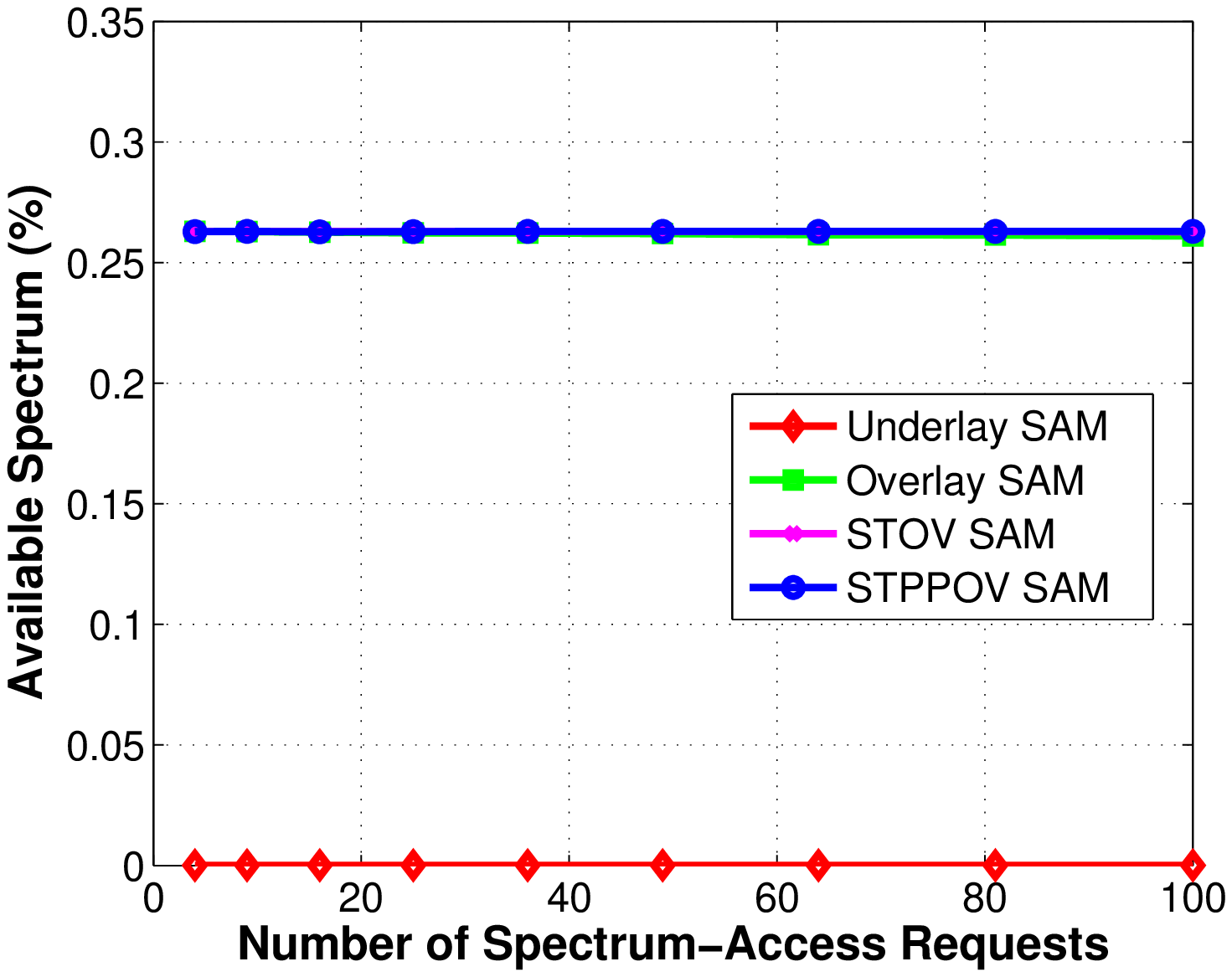}}
{\includegraphics [width=0.464\textwidth, angle=0] {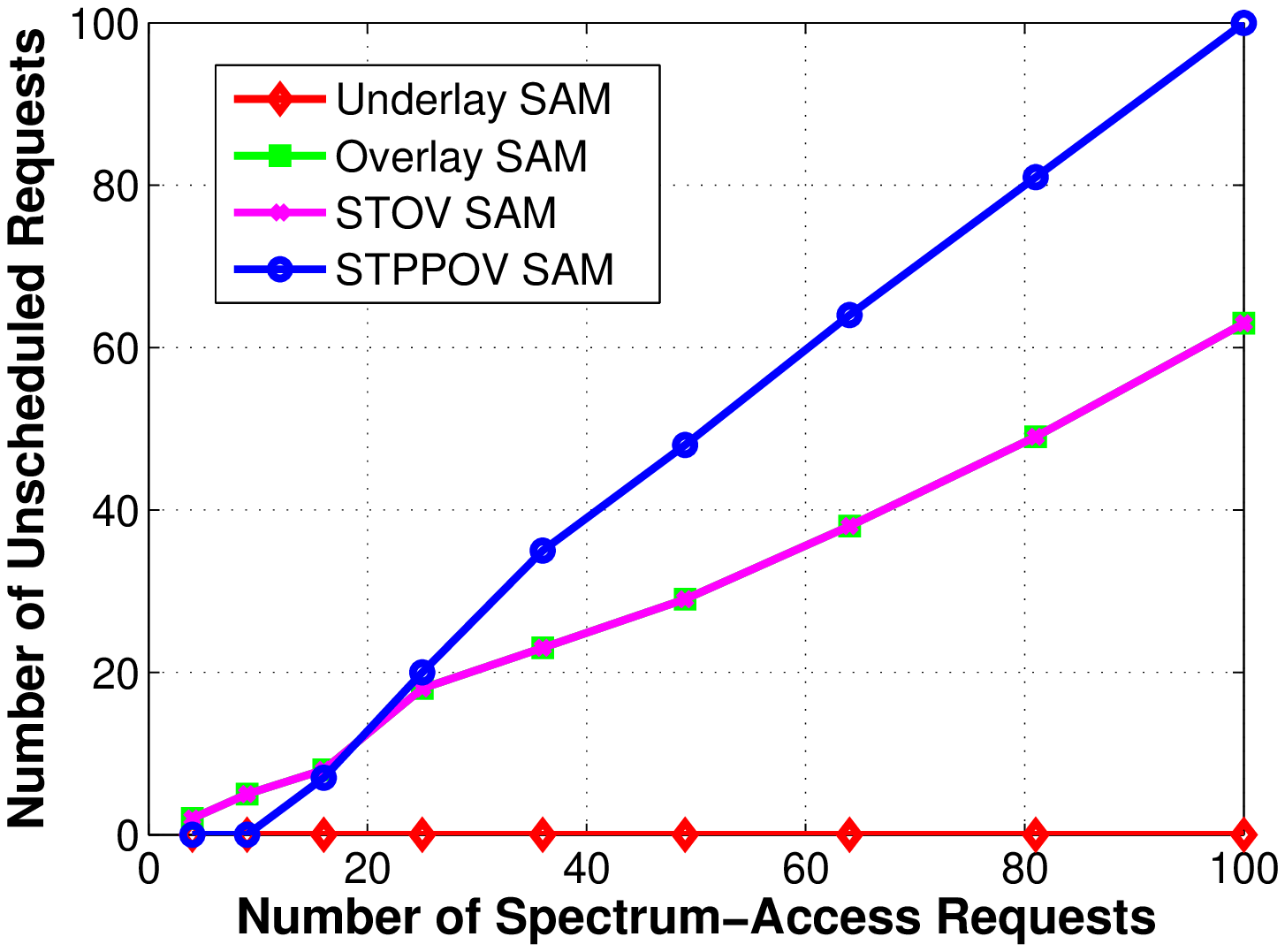}}
\caption{The baseline performance of four SAMs  with varying number of secondary networks when the PU is active \textbf{(The base experiment)}. A SAM with a lower number of unscheduled requests, a lower number of harmfully interfered receivers, and lower spectrum consumption by transceivers (that is, higher amount of available spectrum) makes more efficient use of the available spectrum.}
\label{fig:LL501_ST0}
\end{figure}

The performance of four SAMs for this base setup is shown in Figure \ref{fig:LL501_ST0}. We make following observations about  performance of the SAMs.
\begin{itemize}
	\item In case of Underlay SAM, all the SU networks are exercising spectrum-access, thus the number of scheduled requests is same as the number of SU networks. However, since the transmit-power is heavily constrained,  the signal power at the secondary receivers is very low. Also, these receivers experience interference from the PU transmitter and the transmitters of the cochannel SU networks resulting in unsuccessful reception in most cases.
	\item In case of Overlay SAM, when PU transmitter is not within the sensing range of the SU network transceivers, a spectrum-access is exercised. The SU sensitivity is considered to be -80 dBm translates to 1390 $m$ of sensing range considering the mean PLE of 3.5. Thus, very often, a spectrum-access is not performed due to constraints imposed by the Overlay spectrum-access policy and the number of scheduled requests is poor when PU is present. When a SU transmitter can exercise access to the spectrum, the signal power at the SU receivers may not be high due to constraints on the SU transmit-power and those receivers need to tolerate interference from the cochannel PU and SU transmissions and can result into a significant number of harmfully interfered receivers.
	\item The STOV SAM with unconstrained SU transmit power faces similar issues and shows low performance when PU is present. Because it employs high transmit power, the SINR at the SU receivers is high and receiver spectrum consumption is low and the available spectrum remains high.
	\item The STPPOV SAM shows better performance as compared to Overlay SAM and STOV SAM in the cases SUs can detect the presence of the PU transmitter but their transmissions do not cause harmful interference to the PU receivers. However, the number of  scheduled requests is much lower when available spectrum space is low due to the guard margin enforced to protect the PU receivers from the aggregate interference effect.
\end{itemize}

We note that the four SAMs do not provide protection to SU receivers from harmful interference from cochannel SU transmitters. \textit{The lack of \textbf{network coexistence with non-harmful interference} could be considered as a serious limitation factor in actual practice when the secondary spectrum access is employed for services requiring good link quality}.

\subsubsection{A Case for Higher PU Transmit-Power and Interference-Tolerant Transceivers (Experiment-1)}

In this experiment, we consider a scenario wherein the secondary spectrum access in a licensed band is managed by the spectrum-access policy defined by the incumbent owners of the spectrum. We term this scenario as \textit{Primary Owned Secondary Spectrum Access (POSSA)}. \textbf{POSSA allows improvement in the available spectrum by employing higher PU transmit-power}. We also attempt to improve the available spectrum space by using better quality directional PU receivers with 10 dB minimum desired SINR instead of 20 dB. We consider the SU transceivers to employ directional antennas. 

\begin{figure}[htbp!]
\centering
{\includegraphics [width=0.464\textwidth, angle=0] {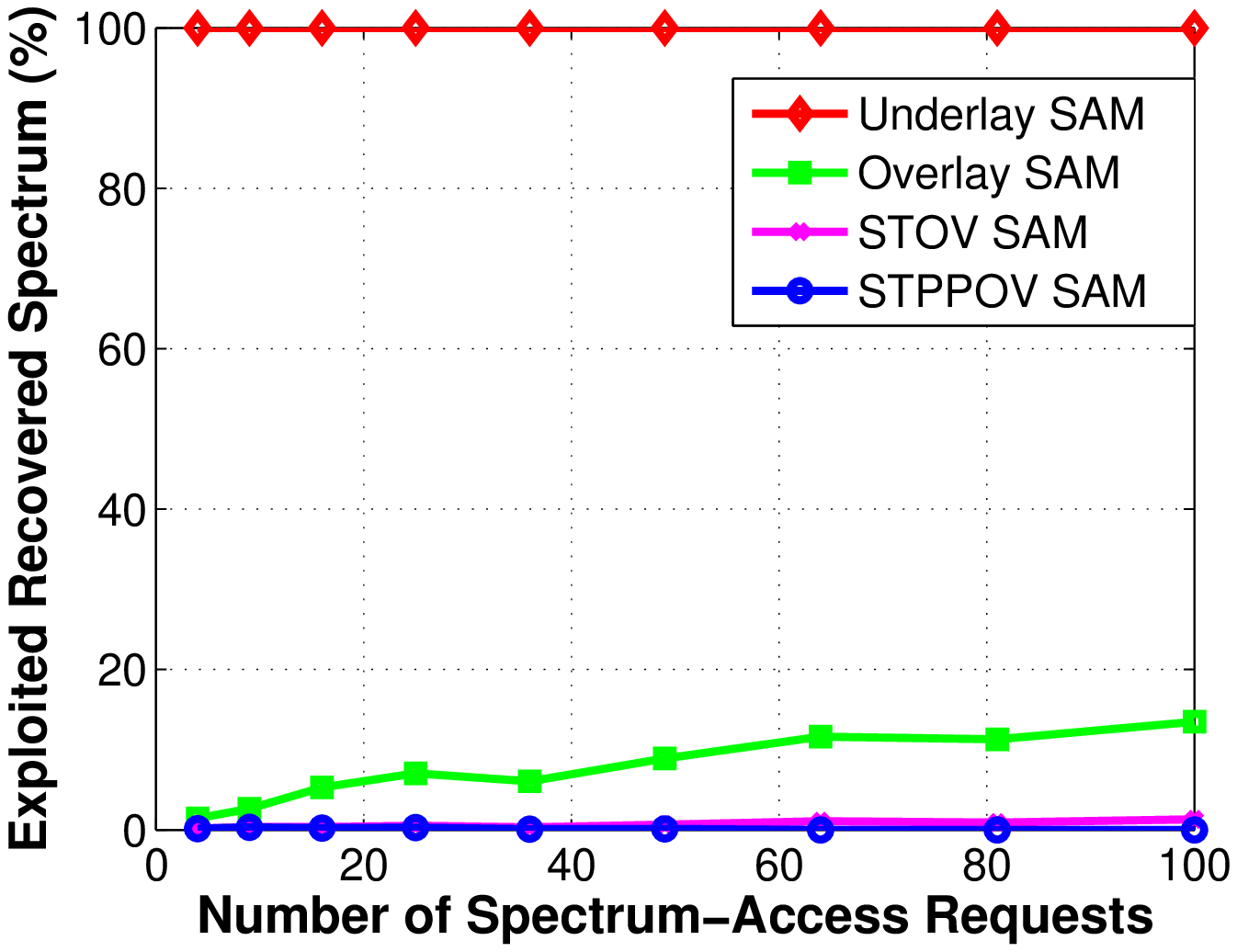}}
{\includegraphics [width=0.464\textwidth, angle=0] {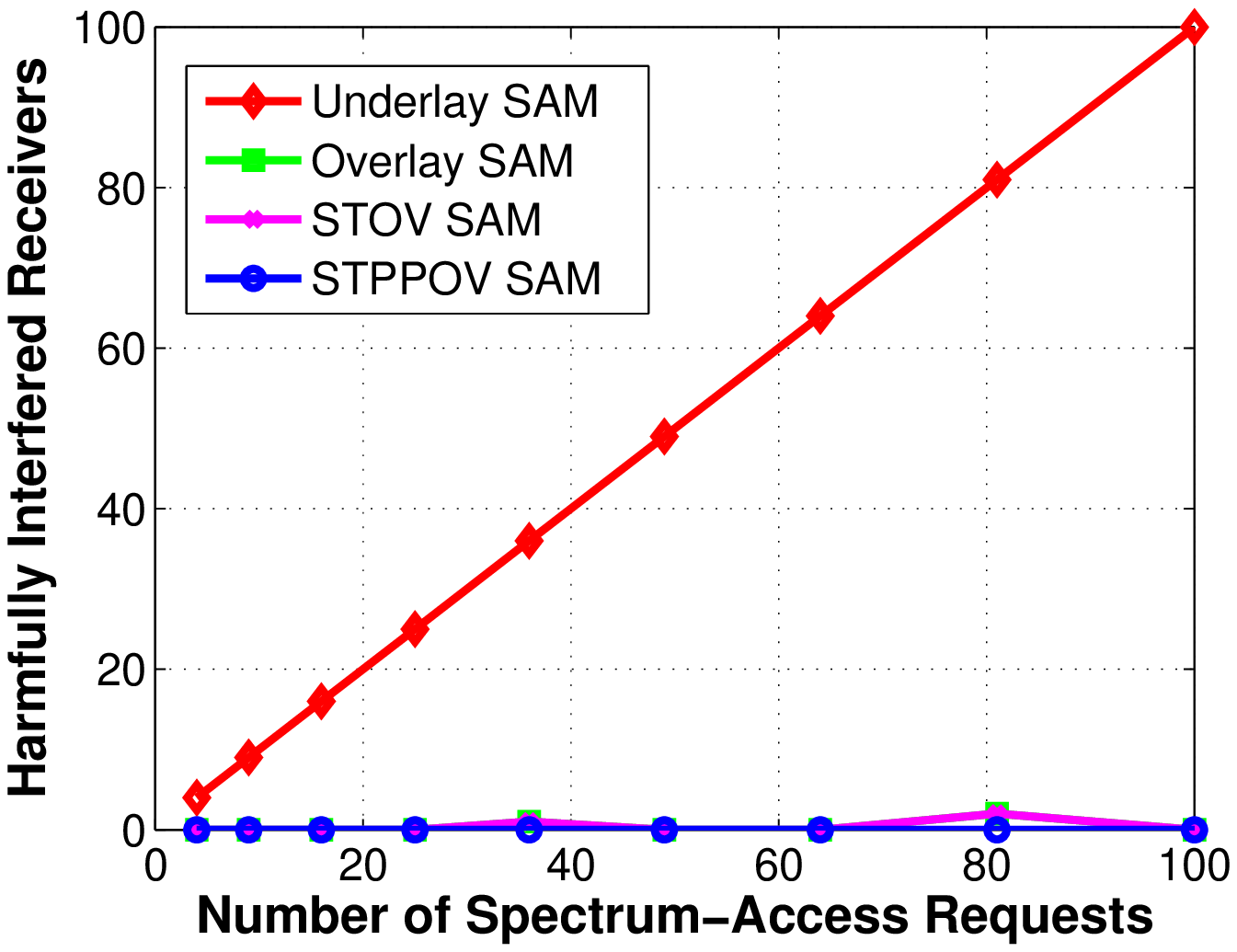}}
{\includegraphics [width=0.464\textwidth, angle=0] {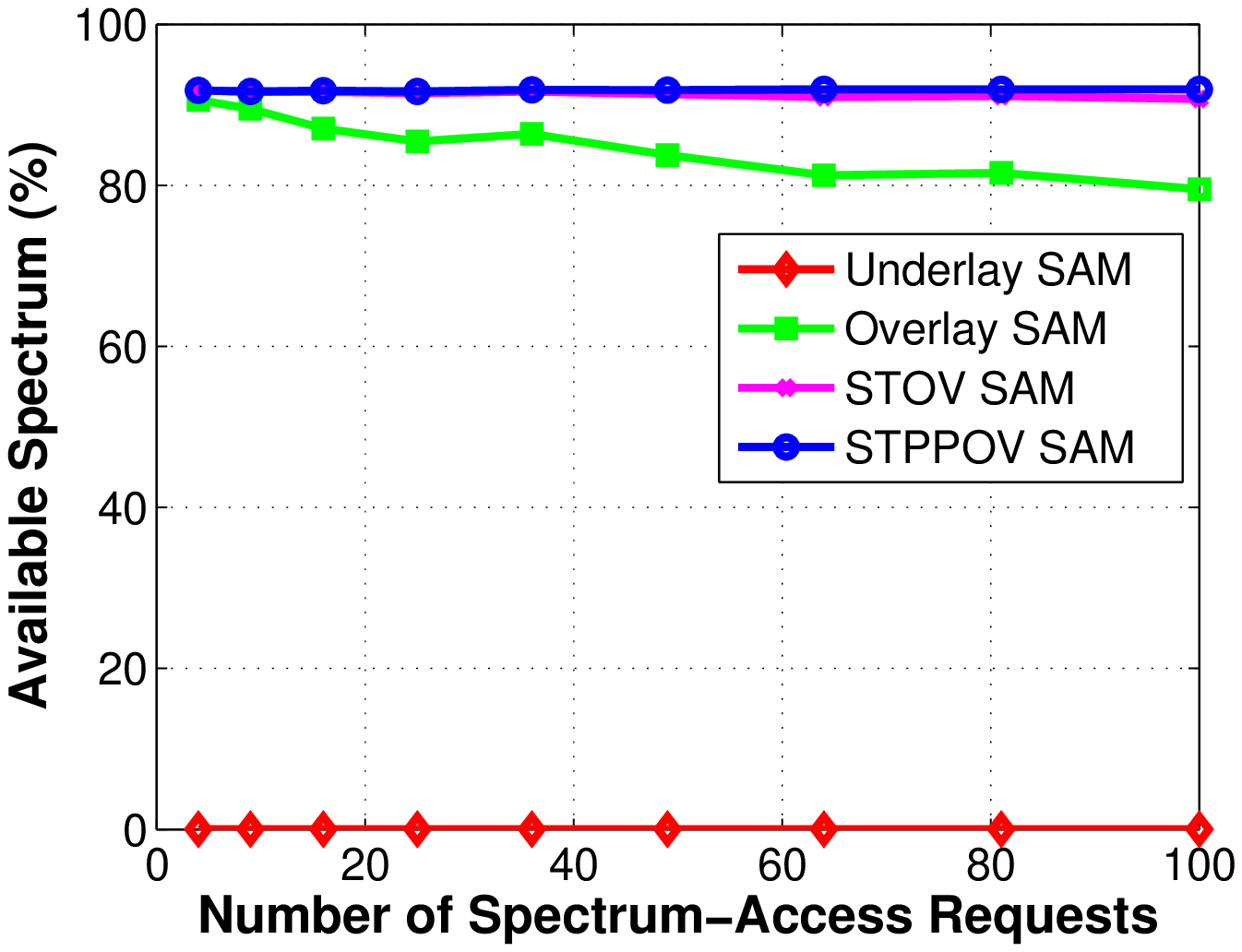}}
{\includegraphics [width=0.464\textwidth, angle=0] {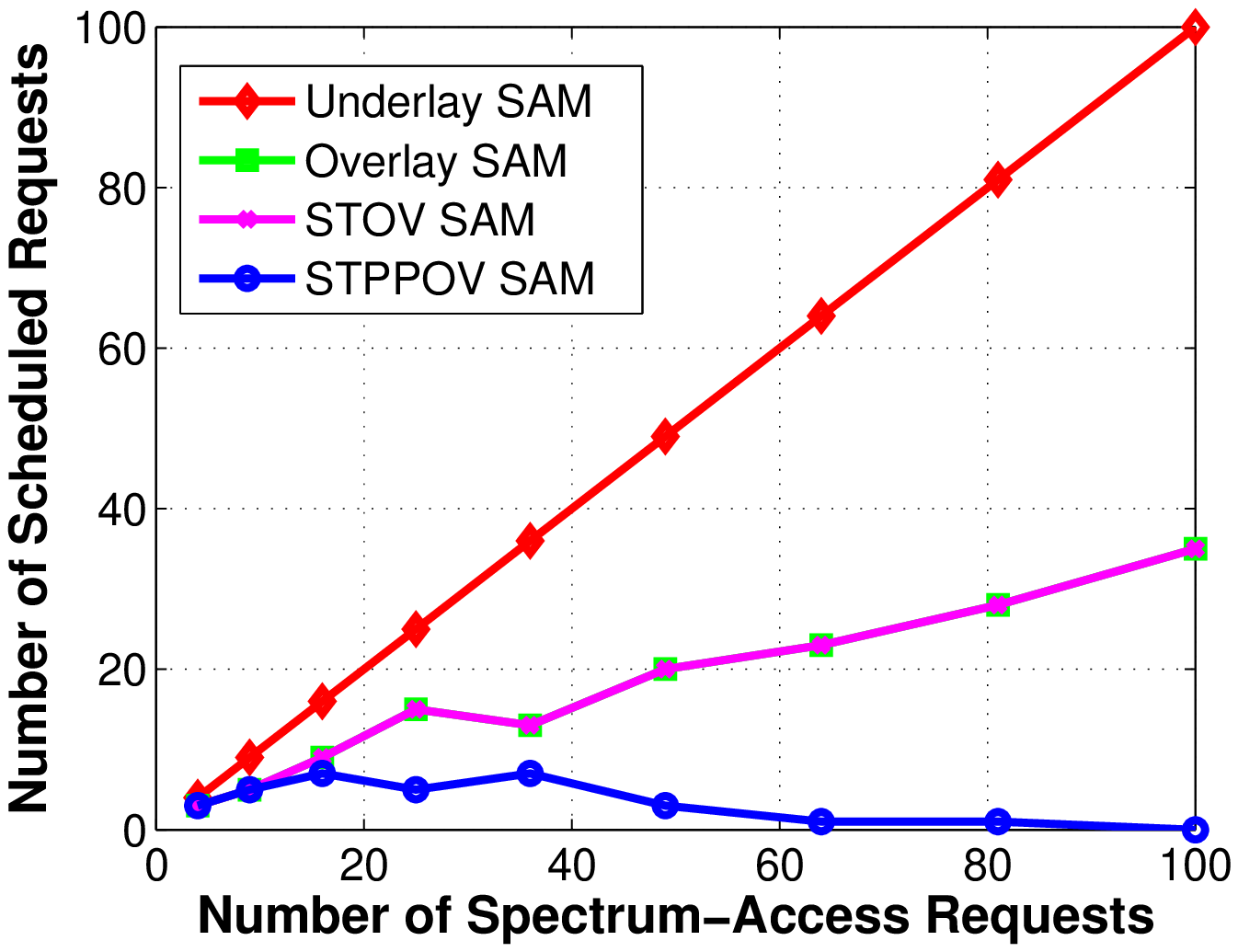}}
\caption{Performance comparison of the four SAMs with varying number of secondary networks when the PU is active \textbf{(Experiment-1)}. In this experiment, the SU transceivers employ directional antennas and the incumbent network is assumed to be using higher transmit power to boost the SINR at the PU receivers in addition to the assumptions from the base-case experiment.}
\label{fig:LL501_ST6}
\end{figure}

From Figure \ref{fig:LL501_ST6}, we observe that the available spectrum has increased from close to $3 \%$ to close to $100 \%$. This is because the interference margin implied by the SINR at the PU receiver is much higher with increase in the PU signal power\footnote{With directional antennas only, the available spectrum increases from close to $3 \%$ to close to $15 \%$. Please refer to Experiment 2 from \cite{oms3_cf1}.}.  Note that Figure \ref{fig:LL501_ST6} shows the number of \textit{scheduled} connections instead of the number of \textit{unscheduled} connections to emphasize the poor numbers of the scheduled connections even when the available spectrum is close to $100\%$.

\textbf{\textit{The secondary access scenario is not encouraging.}} The performance of STPPOV SAM which ensures protection of PU receivers while opportunistically improving the throughput for secondary networks is found to be \textit{not} able to exploit a substantial portion of the available spectrum even when nearly  $100\%$ of the spectrum is available. 

The inability of the SAMs to exploit the available spectrum is due to \textit{inefficient scheduling} of spectrum-access requests. The cochannel networks are spatially overlapping. With a greedy strategy and a limit on the maximum transmit-power, often many of the spatially overlapping spectrum-access requests are unable to get scheduled. In this experiment, we focus on improving the spectrum sharing performance by choosing an alternate scheduling approach. 

\subsubsection{Favoring the Requests with Lower Spectrum Consumption Cost (Experiment-2)}
Under the QDSA paradigm, the role of SAM can be viewed as allocating a quantified amount of spectrum to networks (i.e. network transmitters and receivers). \textit{To accommodate a large number of spectrum-access requests, a SAM needs to define a spectrum-access policy for a spectrum-access request in such a fashion that the spectrum consumed by its transceivers is minimized and in turn the available spectrum is maximized.} Thus, our objective for designing a spectrum allocation algorithm is to maximize the number of scheduled connections and the available spectrum subject to the constraint that no (primary or secondary) receiver experiences SINR lower than the receiver-specific minimum desired SINR.

In this experiment, we add the `NSC-CX SAM', that schedules spectrum-access requests based on the \textit{minimal network spectrum consumption} and emphasizes coexistence\footnote{Here, we assume the networks are \textit{coexisting} when none of the network-receivers are harmfully interfered.} between all cochannel networks. The steps are described previously in Section 4.B.3. Here, we note that the spectrum-access requests employing
\begin{itemize}
	\item a smaller network-range are favored by the `NSC-CX SAM' as the receivers experience higher SINR and the network spectrum consumption weight is lower,
	\item directionality are favored as it helps to improve SINR for all receivers and reduces the network spectrum consumption.
\end{itemize}
\begin{figure}[htbp!]
\centering
{\includegraphics [width=0.464\textwidth, angle=0] {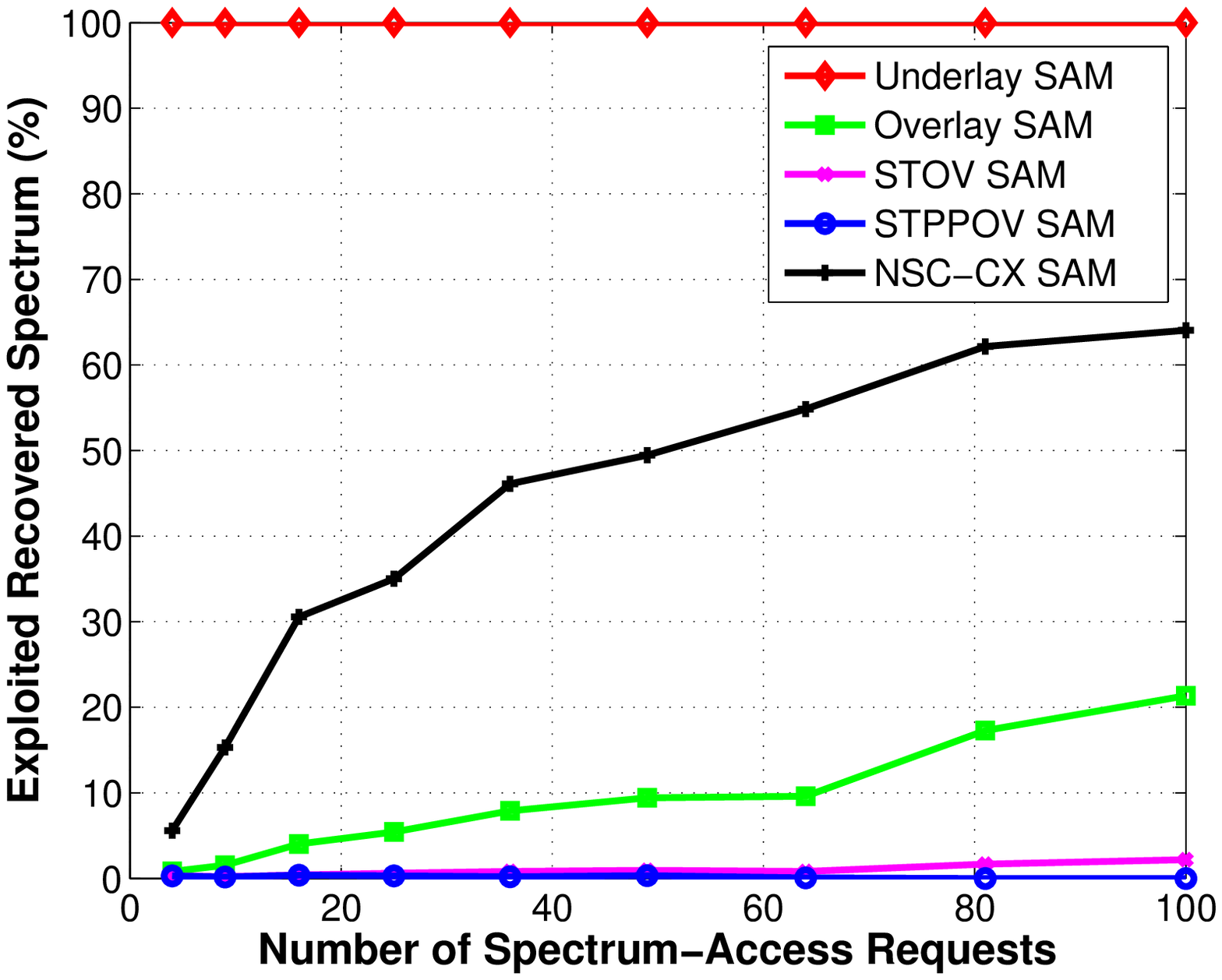}}
{\includegraphics [width=0.464\textwidth, angle=0] {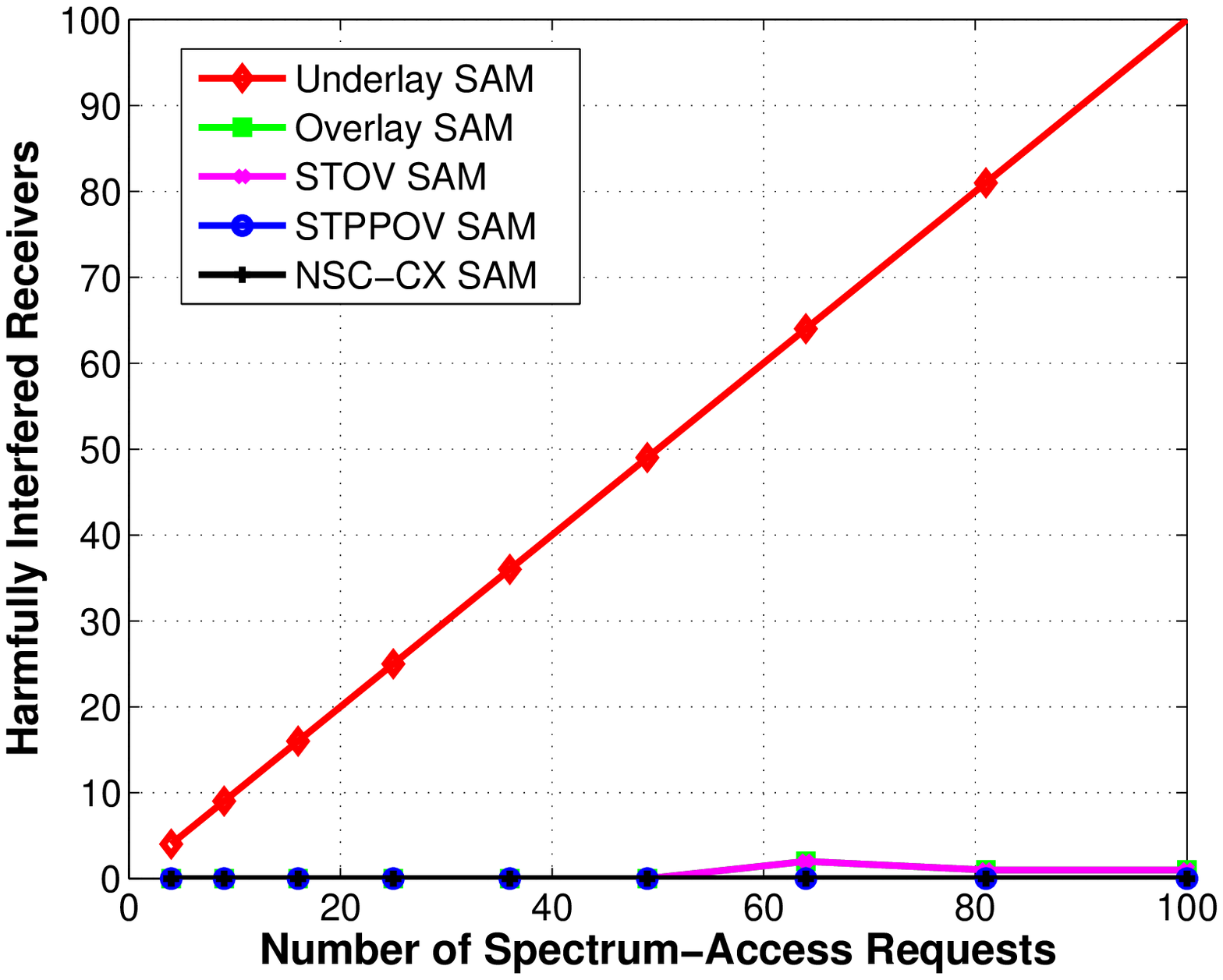}}
{\includegraphics [width=0.464\textwidth, angle=0] {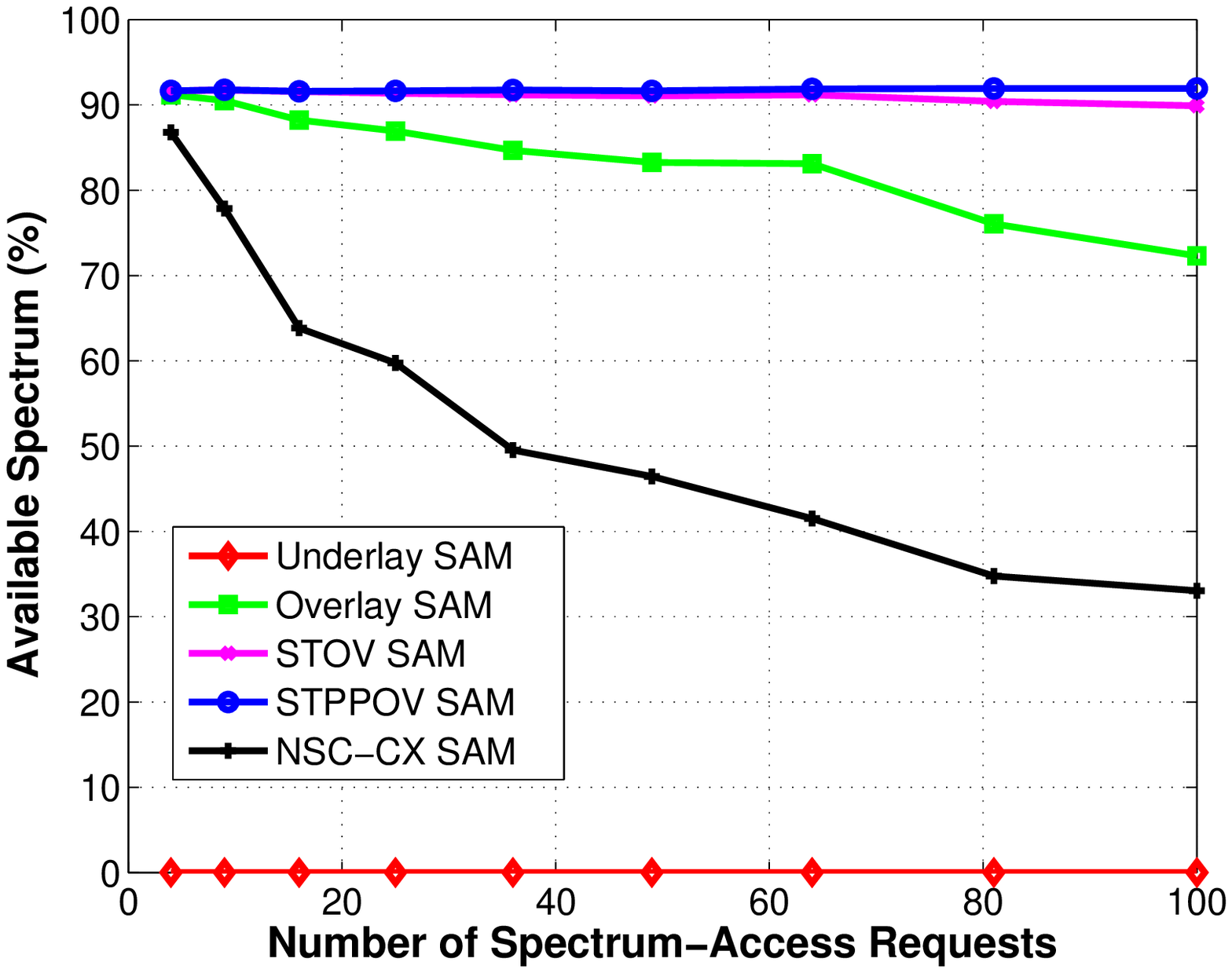}}
{\includegraphics [width=0.464\textwidth, angle=0] {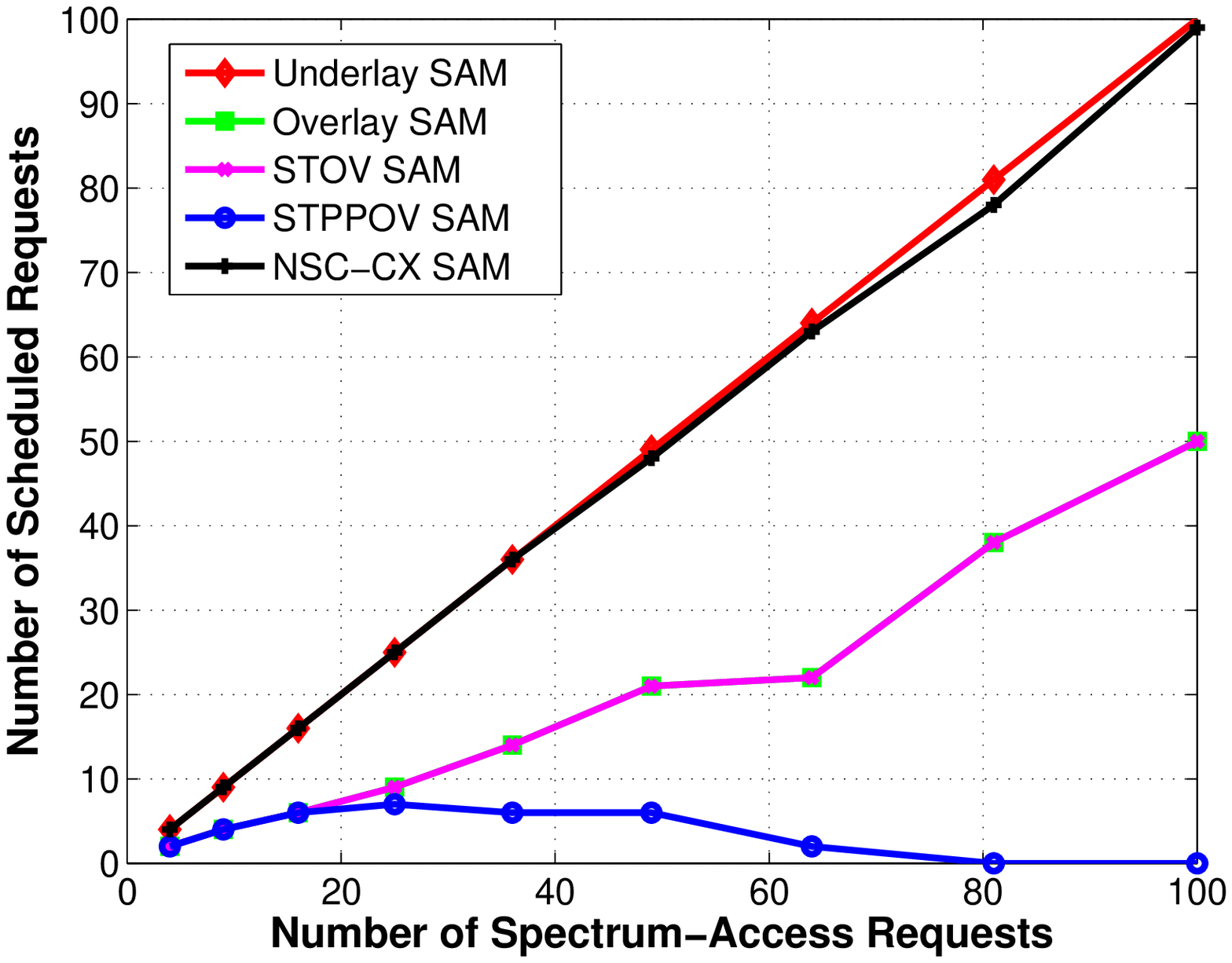}}
\caption{Performance comparison of SAMs with varying number of secondary networks when the PU is active  \textbf{(Experiment-2)}. The `NSC-CX SAM' favors spectrum-access requests with lower network spectrum consumption and also ensures coexistence with cochannel primary and secondary networks.}
\label{fig:LL501i_ST6}
\end{figure}
The setup for the experiment is kept the same as Experiment-1. The performance of the five SAMs is shown in Figure \ref{fig:LL501i_ST6}. The performance of `NSC-CX SAM' in terms of the number of scheduled requests seems very promising as most of the secondary spectrum-access requests are serviced. We note a salient feature of the algorithm is that it allows each network to perform the best in terms of minimizing the spectrum consumption \cite{oms2_sca}. This makes it possible to independently consider each request and then define the order for scheduling. A few more observations on the performance of `NSC-CX SAM':
\begin{itemize}
	 \item As mentioned earlier the objective of `NSC-CX SAM' is to minimize network spectrum consumption and maximize the number of scheduled connections. Thus, it does not have necessarily provide optimal performance in terms of scheduling and power allocation.
	 \item As expected, the number of harmfully interfered receivers with `NSC-CX SAM' is zero.
	 \item The exploited spectrum is higher due to the higher number of scheduled connections and it is the same reason why the available spectrum is lower with `NSC-CX SAM'.   
\end{itemize}

\subsubsection{A Case for the Knowledge of the PU Receiver Positions (Experiment-3)}

The obvious next question is can we do better? In the previous experiments, we had assumed the primary receivers are located at the worst-case positions. This is mainly because, traditionally the primary receivers are assumed to be passive receivers. \textbf{With incumbents playing an active role}, it makes a reasonable case to exploit the knowledge of the \textit{actual} receiver positions to extract more value from the existing spectrum.  Without the knowledge of the receiver positions, the secondary spectrum access is only possible \textit{outside} the range of the primary service network and this is not as appealing from a business perspective. 

In this experiment, we position the primary receivers at half the range, that is, 250 $m$. We assume the knowledge of the actual receiver positions is available in case of the `STPPOV SAM' and `NSC-CX SAM'.
\begin{figure}[htbp!]
\centering
{\includegraphics [width=0.464\textwidth, angle=0] {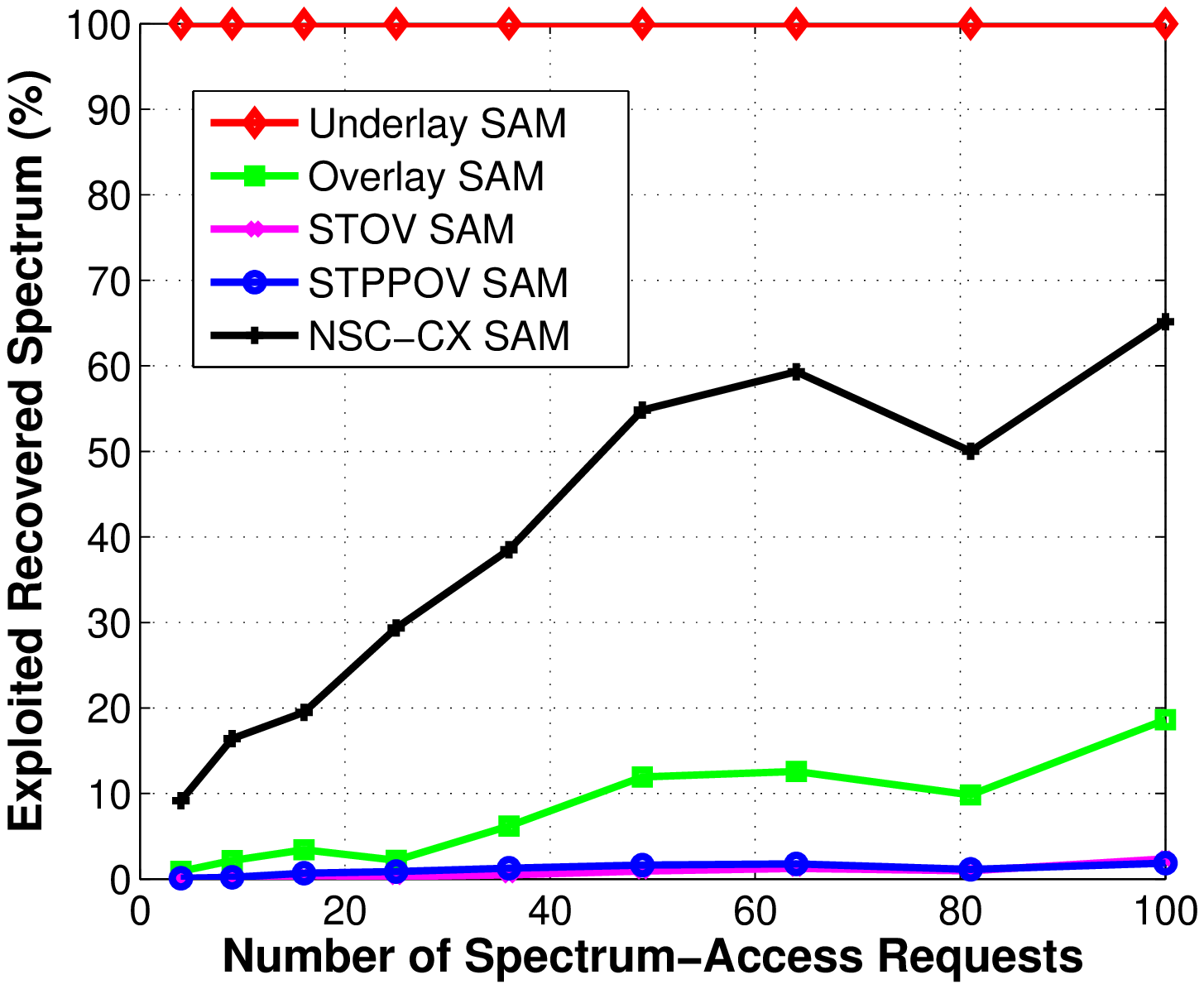}}
{\includegraphics [width=0.464\textwidth, angle=0] {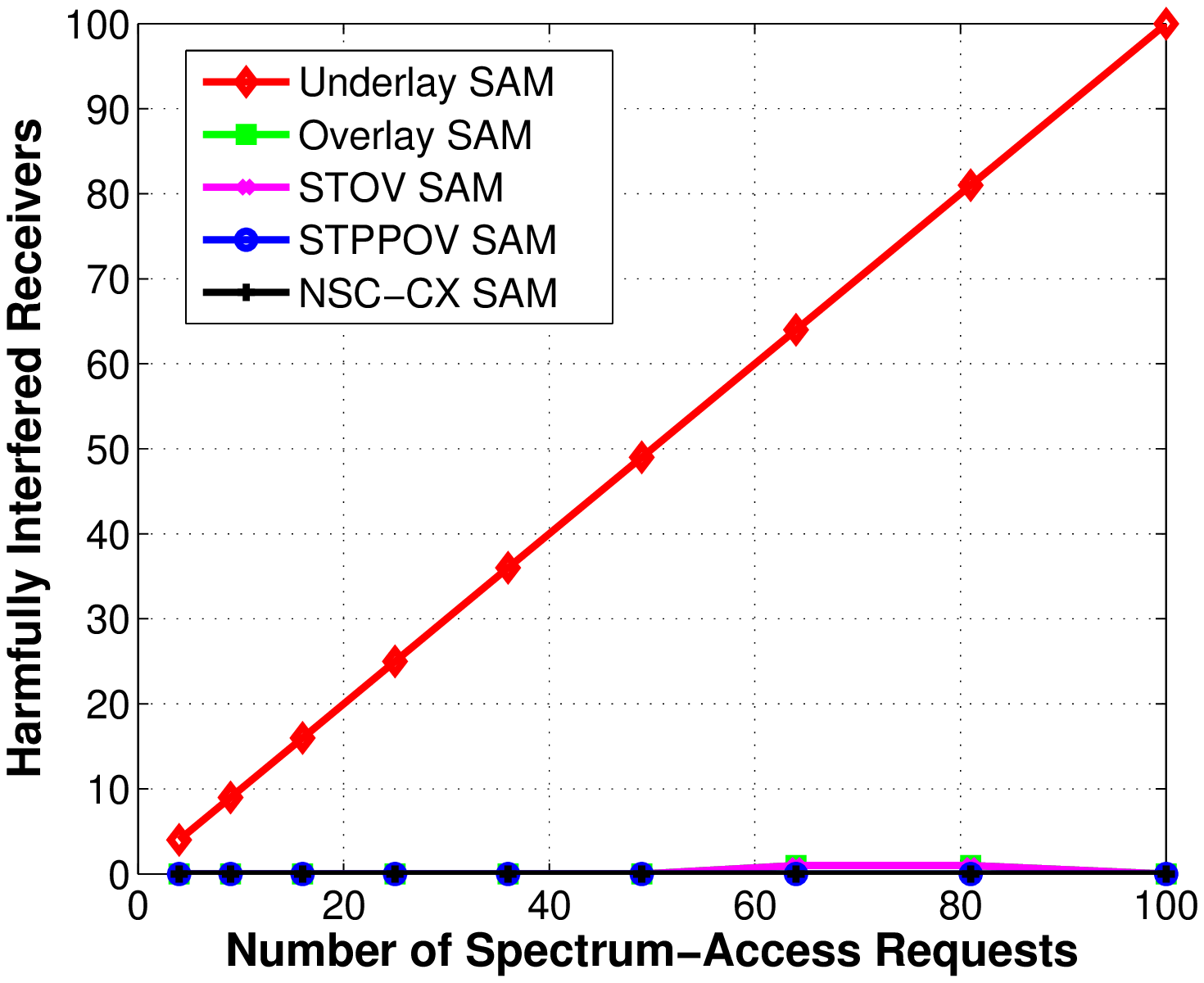}}
{\includegraphics [width=0.464\textwidth, angle=0] {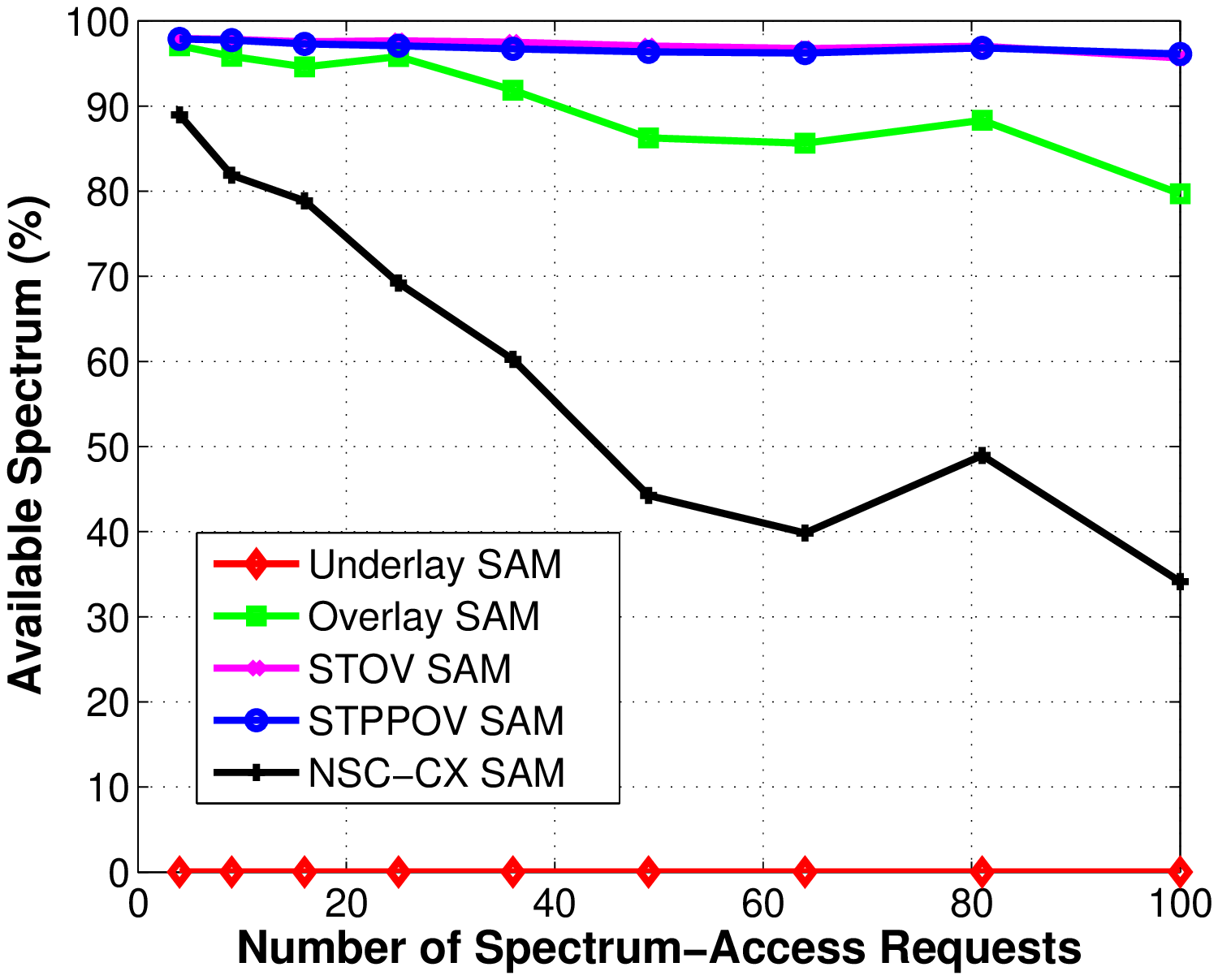}}
{\includegraphics [width=0.464\textwidth, angle=0] {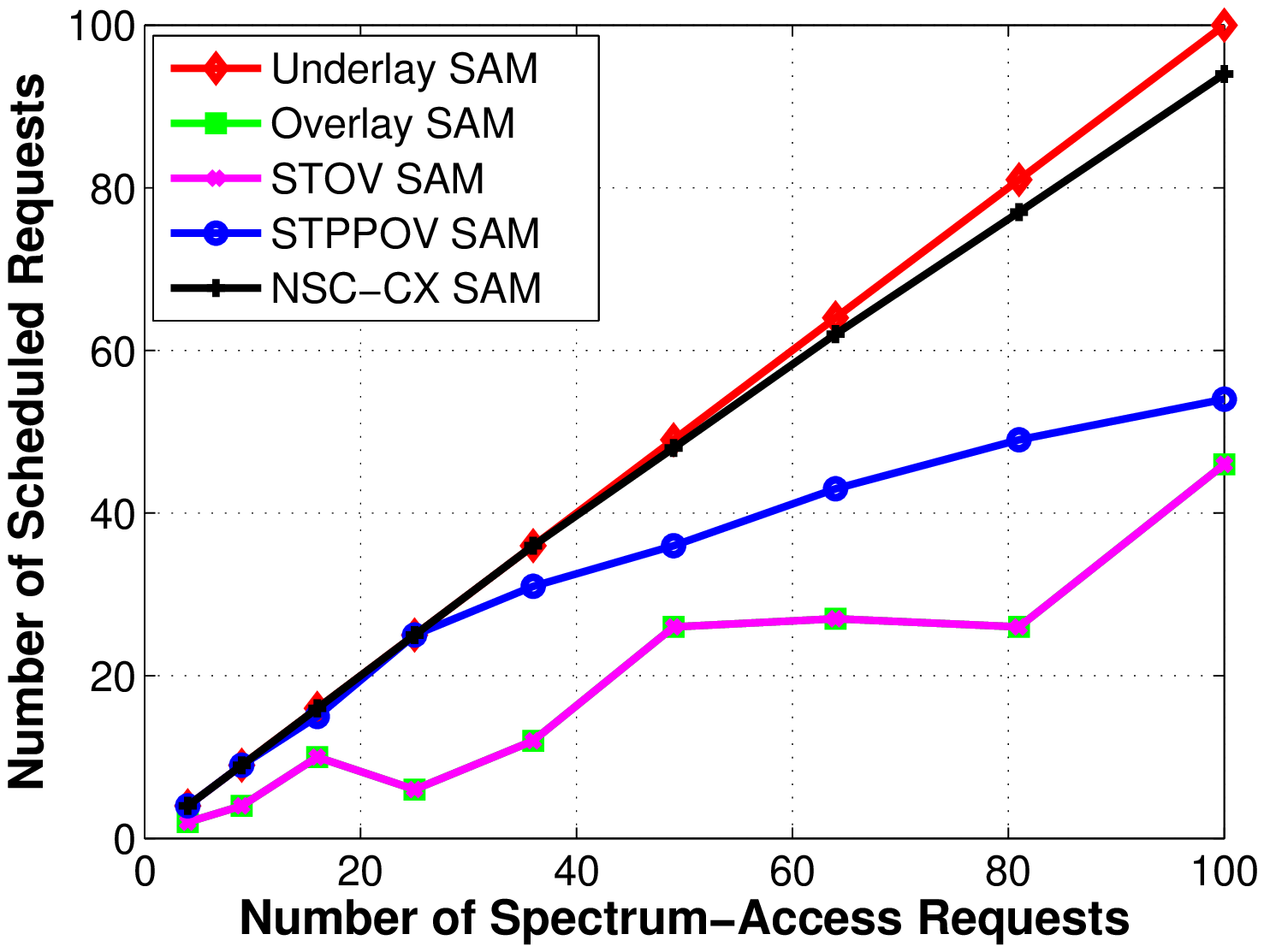}}
\caption{Performance comparison of the five SAMs with the varying number of secondary networks when the PU is active \textbf{(Experiment-3)}. In this experiment, the primary receivers are not positioned at the worst case locations but at a distance of half the range of the service networks. It results in a larger exploitable available spectrum thereby improving SAM performance.}
\label{fig:LL501_ST7}
\end{figure}
From Figure \ref{fig:LL501_ST7}, we observe that the performance of the `STPPOV SAM'  has improved and the number of scheduled requests by `STPPOV SAM' is higher than the Overlay SAM and STOV SAM. As the PU receivers are positioned at a distance of half the range of the service networks, the \textit{exploitable} available spectrum in case of `STPPOV SAM' has increased. 

The number of scheduled requests for `NSC-CX SAM' has also slightly improved because the networks within the primary service coverage area, that are not close to any of the primary receivers, are able to exercise the secondary spectrum access.

\subsubsection{The Case for Small-Cell Secondary Networks (Experiment-4)}

\begin{figure}[htbp!]
\centering
{\includegraphics [width=0.464\textwidth, angle=0] {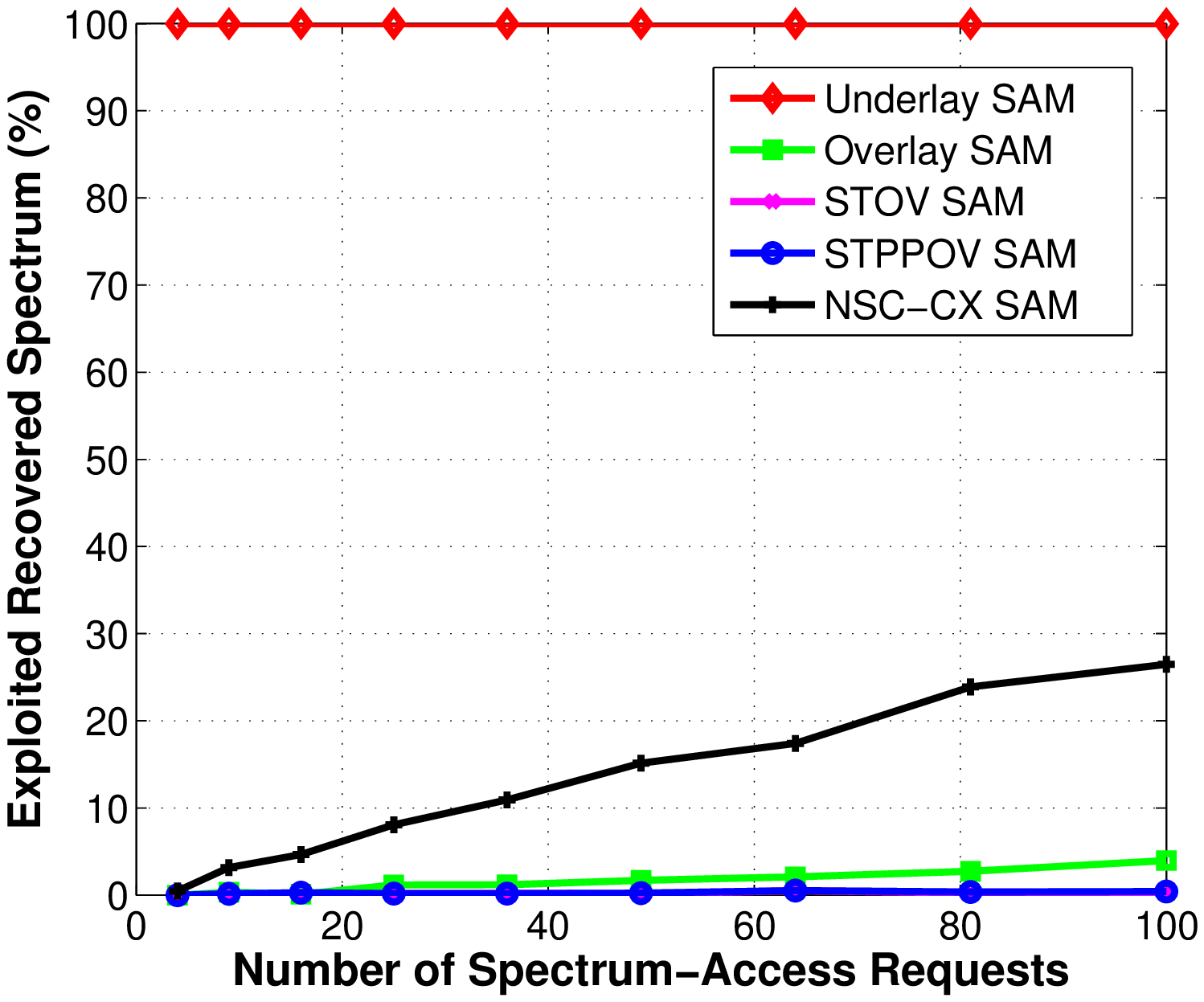}}
{\includegraphics [width=0.464\textwidth, angle=0] {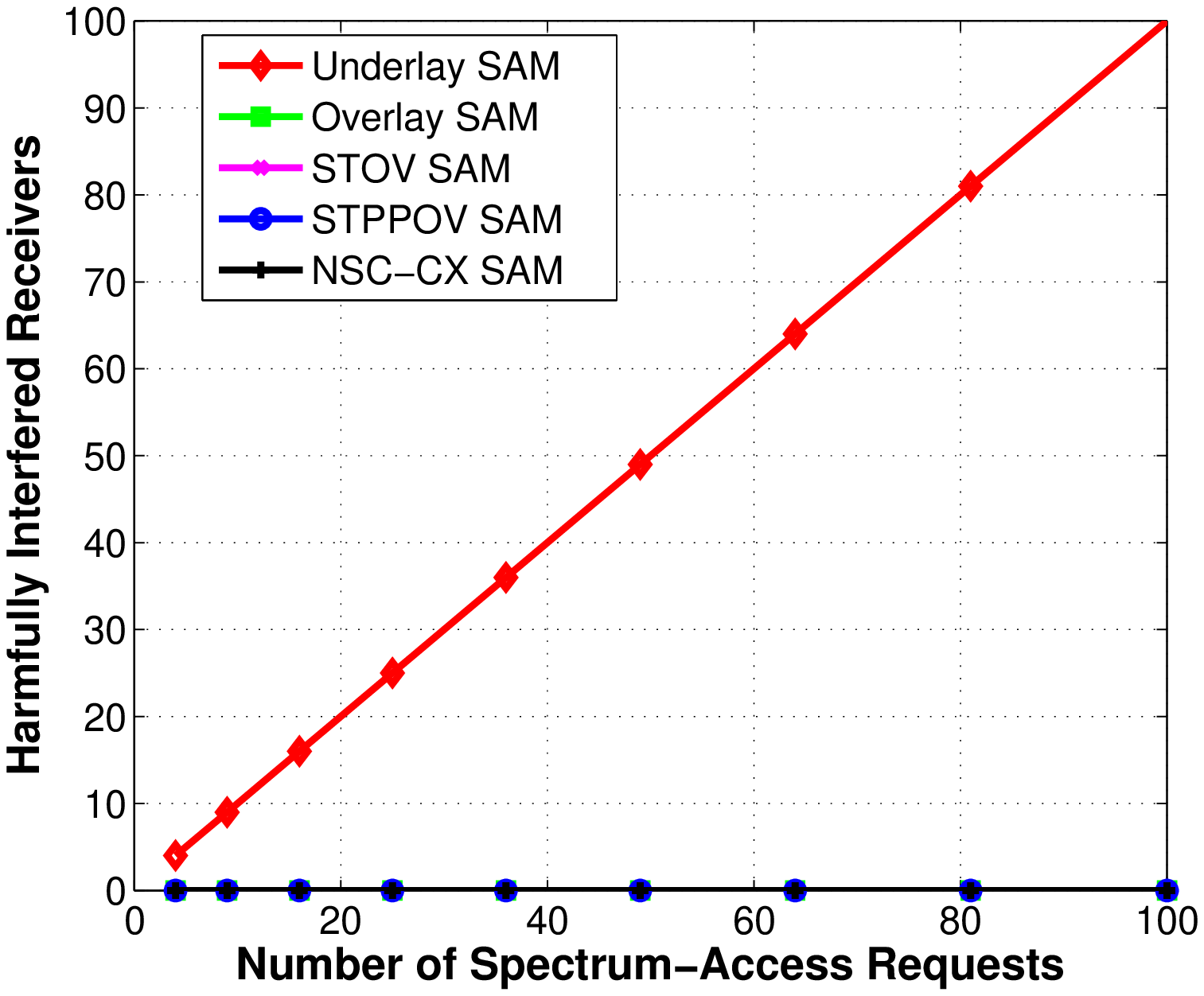}}
{\includegraphics [width=0.464\textwidth, angle=0] {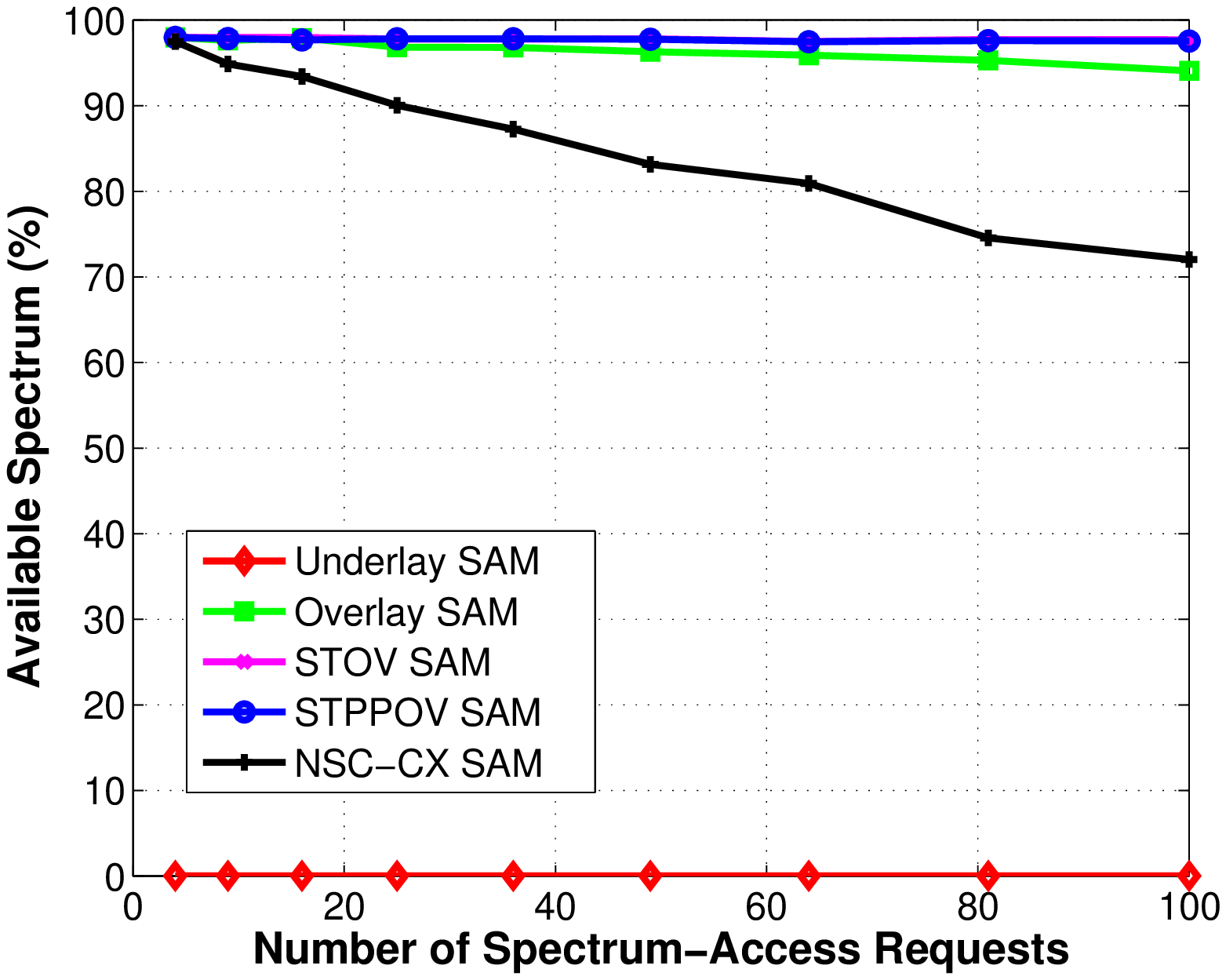}}
{\includegraphics [width=0.464\textwidth, angle=0] {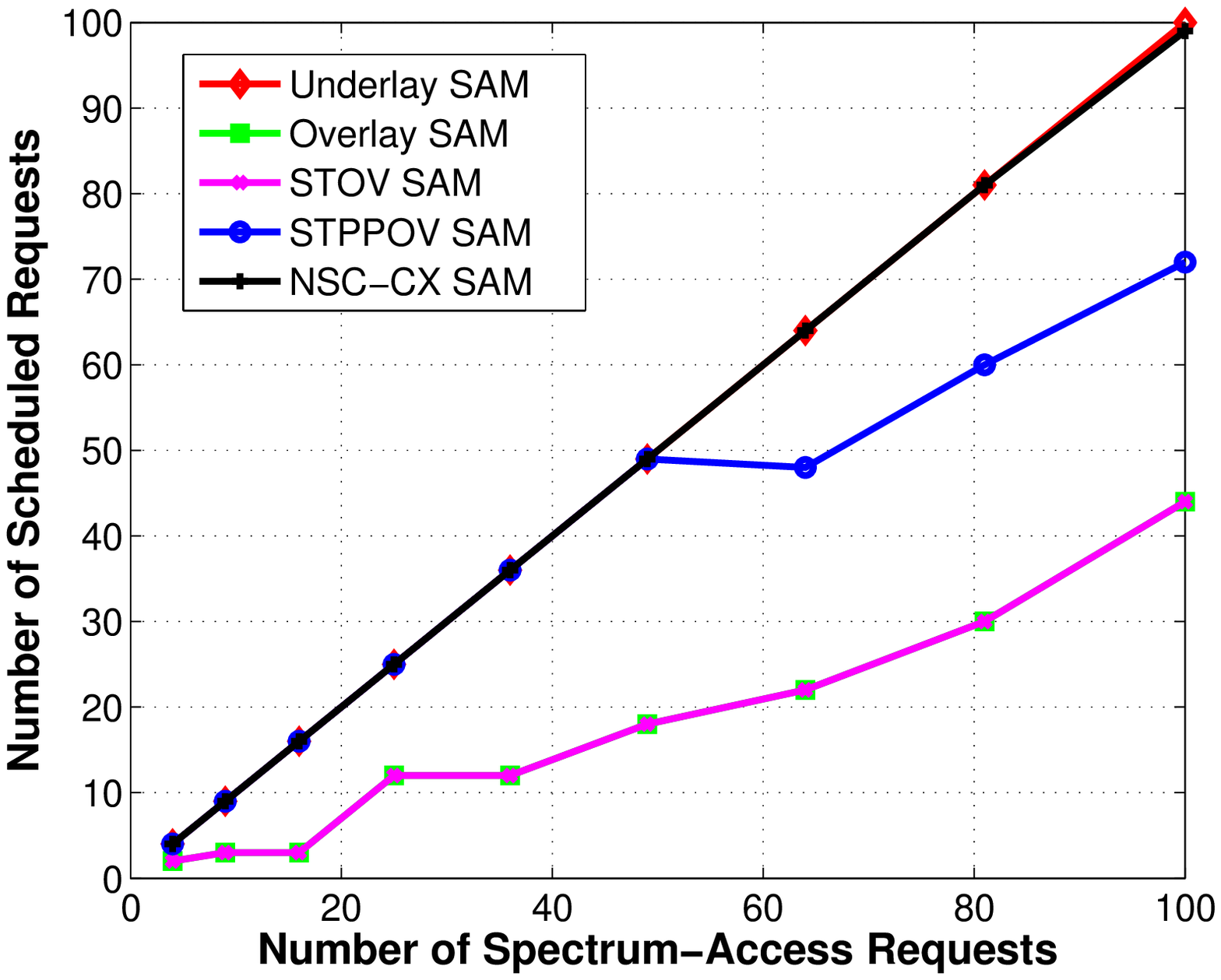}}
\caption{The performance comparison of SAMs with the varying number of the secondary networks when the PU is active \textbf{(Experiment-4)}. In this experiment, the secondary networks are assumed to exercise spectrum-access with a small network-range. With secondary receivers being closer to the respective secondary transmitters resulting in better SINR and higher number of scheduled spectrum-access requests.}
\label{fig:LL501_ST8}
\end{figure}

The next question is how far could we go with scheduling of the secondary spectrum-access requests? As we can see from the previous experiment, the available spectrum in case of `NSC-CX SAM' is close to $40\%$ when the number of requests is 100. As the number of scheduled connections are increased, the available spectrum may be insufficient. In this experiment, we reduce the network-range of the secondary networks to 40 $m$ and evaluate the performance of the SAMs. From Figure~\ref{fig:LL501_ST8}, we observe that
\begin{itemize}
  \item the exploited spectrum has reduced and the available spectrum has significantly increased. The available spectrum with `STPPOV SAM' is higher than `NSC-CX SAM'. This is due to behavior of `STPPOV SAM' to choose transmit-powers that are \textit{not very high} as is the case with `NSC-CX SAM' thus limiting interference power but \textit{not low enough }to significantly increase the receiver consumed spectrum. The same behavior is used by the `Overlay SAM' and `STOV SAM', and the available spectrum is high in case of those mechanisms as well.
	\item the performance of `NSC-CX SAM' and `STPPOV SAM' in terms of the number of connections has also improved. The higher number of scheduled requests in this experiment setup is attributed to the fine granular spatial spectrum access opportunities getting exercised. \textit{We argue that for realizing the potential of DSA, finer granularity of spatial reuse is essential.}
\end{itemize}

\subsubsection{The Case for Dynamic Spectrum Sharing or Inactive PU (Experiment-5)}

\begin{figure}[htbp!]
\centering
{\includegraphics [width=0.464\textwidth, angle=0] {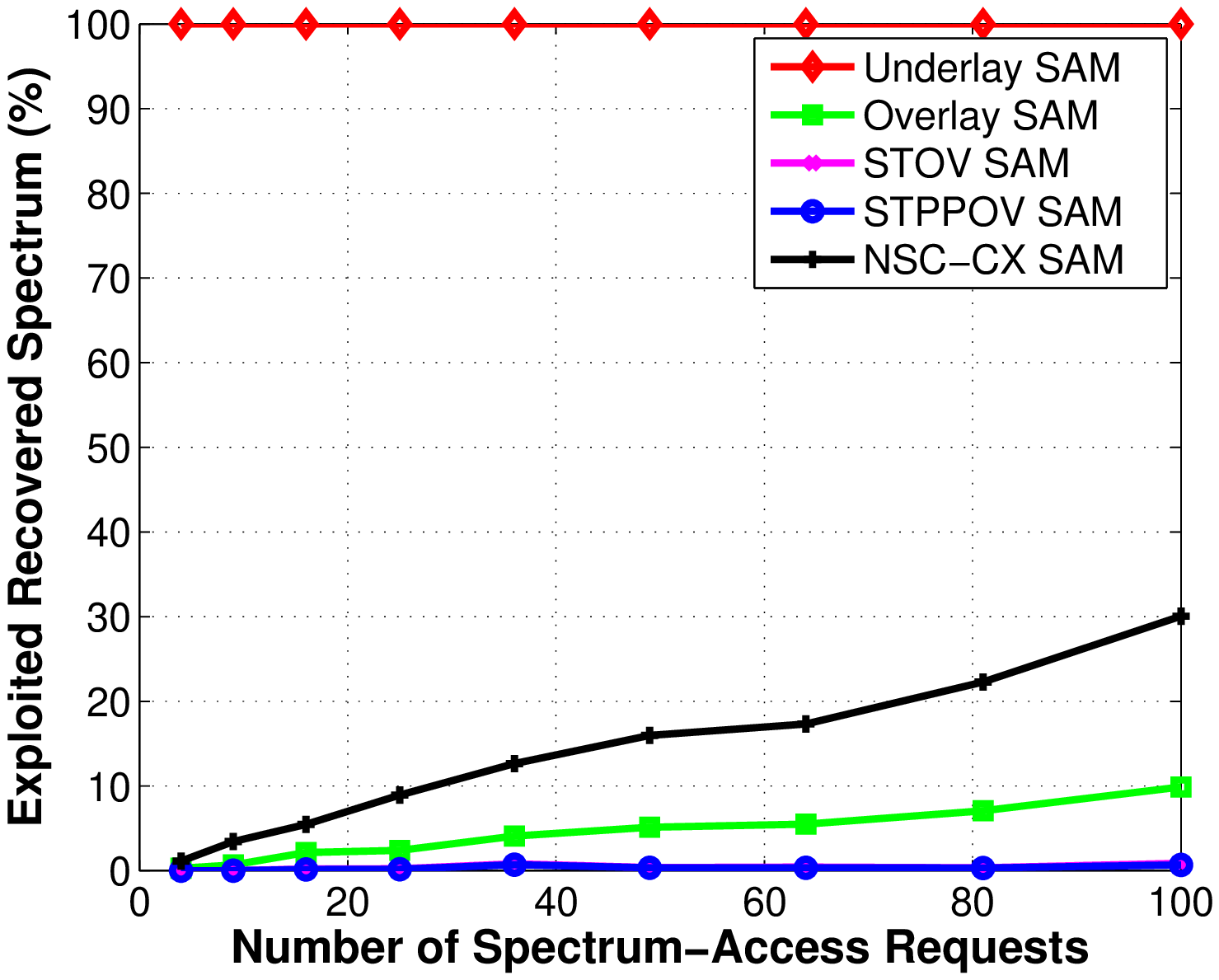}}
{\includegraphics [width=0.464\textwidth, angle=0] {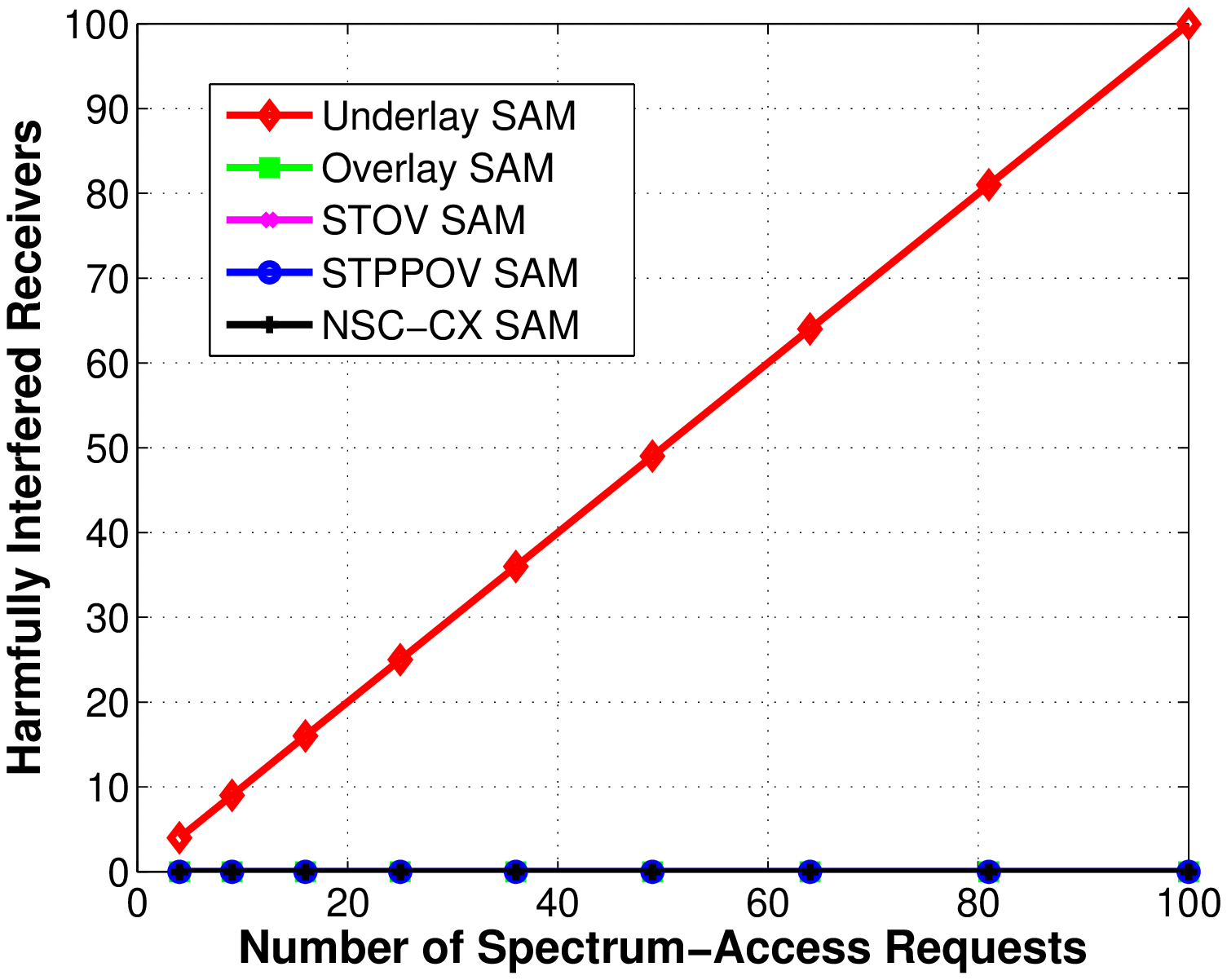}}
{\includegraphics [width=0.464\textwidth, angle=0] {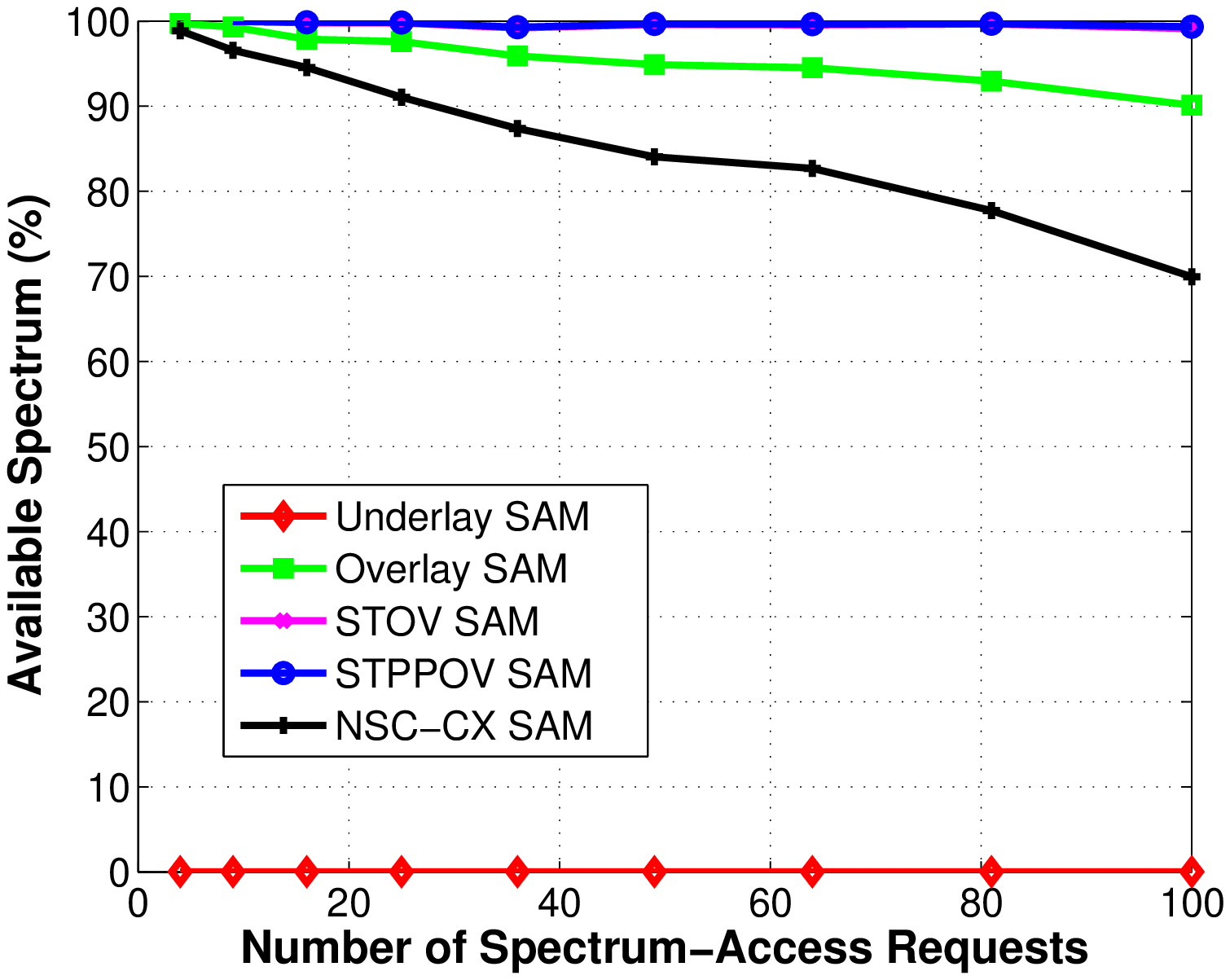}}
{\includegraphics [width=0.464\textwidth, angle=0] {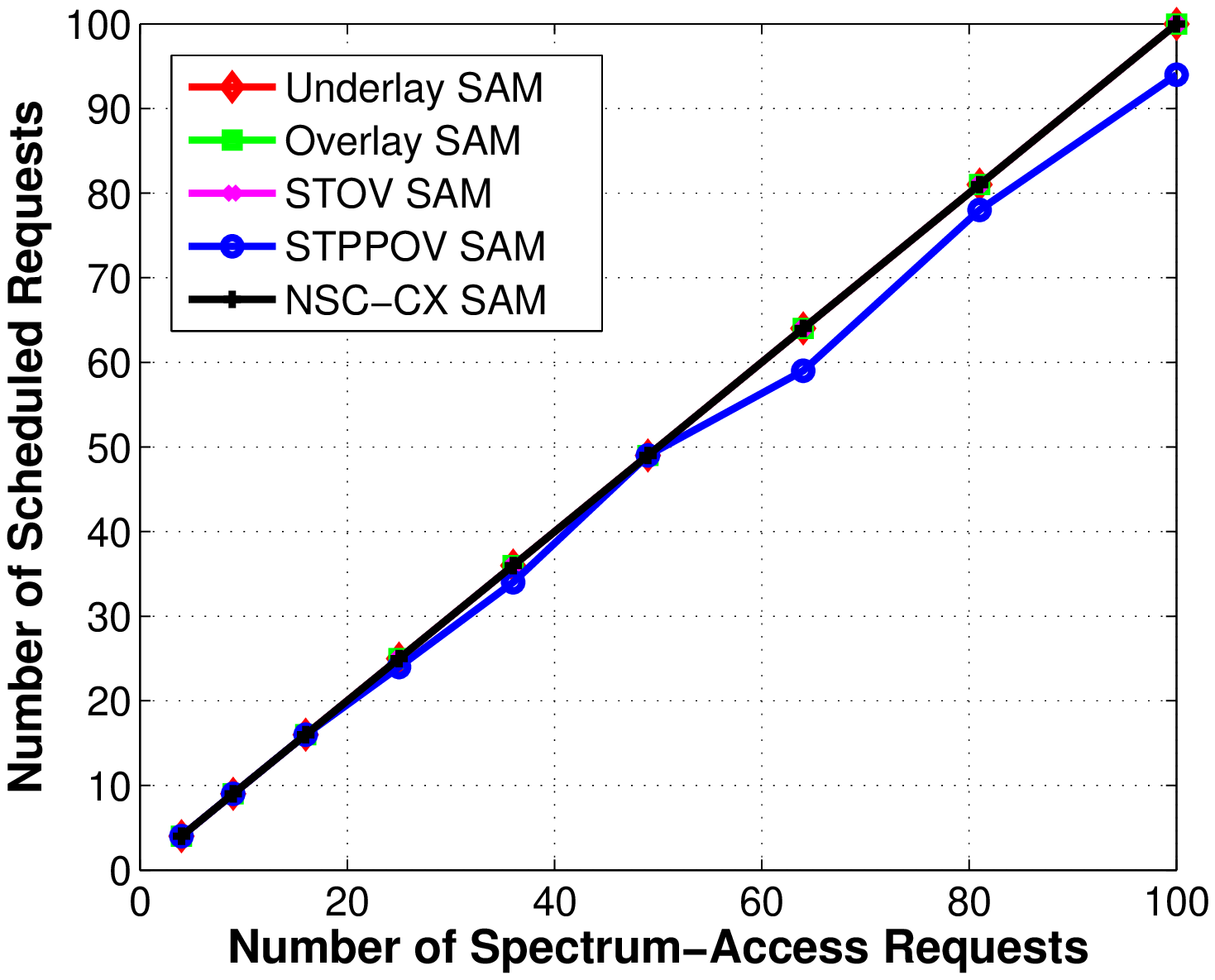}}
\caption{The performance comparison of SAMs with the varying number of the secondary networks \textbf{(Experiment-5)}. In this experiment, the primary network is assumed to be inactive. The performance of all the SAMs is higher with much higher available spectrum space.}
\label{fig:LL501_ST9}
\end{figure}

Now, we consider a more generic case wherein the spectrum rights with all the networks are equal. In the context of POSSA, this case could also be seen the case when the primary service network is \textit{not active}. In this experiment, all the networks are randomly positioned. The range of all the networks is 40 $m$. The minimum desired SINR at the receivers is 3 dB. The transmitters and receivers are directional with $60^{\circ}$ antenna beamwidth. From Figure~\ref{fig:LL501_ST9}, we observe that
\begin{itemize}
	\item Except for `STPPOV SAM', all SAMs have been able to schedule all the spectrum-access requests. The `STPPOV SAM' is not able to support $100 \%$ of the requests because of the guard margin setting to avoid the aggregate interference effect. \textit{This experiment brings out the potential of dynamic spectrum sharing paradigm.}
	\item There are no harmfully interfered receivers for all SAMs other than the Underlay SAM. This is primarily attributed to  the fine granular spatial spectrum access (smaller ranges of the networks). 
\end{itemize}
It may not be always possible to choose a smaller network range depending on the nature of the desired service. In the next experiment, we characterize the behavior of the SAMs based on the network range.

\subsubsection{Characterizing the Effect of Network Range in Open Dynamic Spectrum Sharing Model (Experiment-6)}

\begin{figure}[htbp!]
\centering
{\includegraphics [width=0.454\textwidth, angle=0] {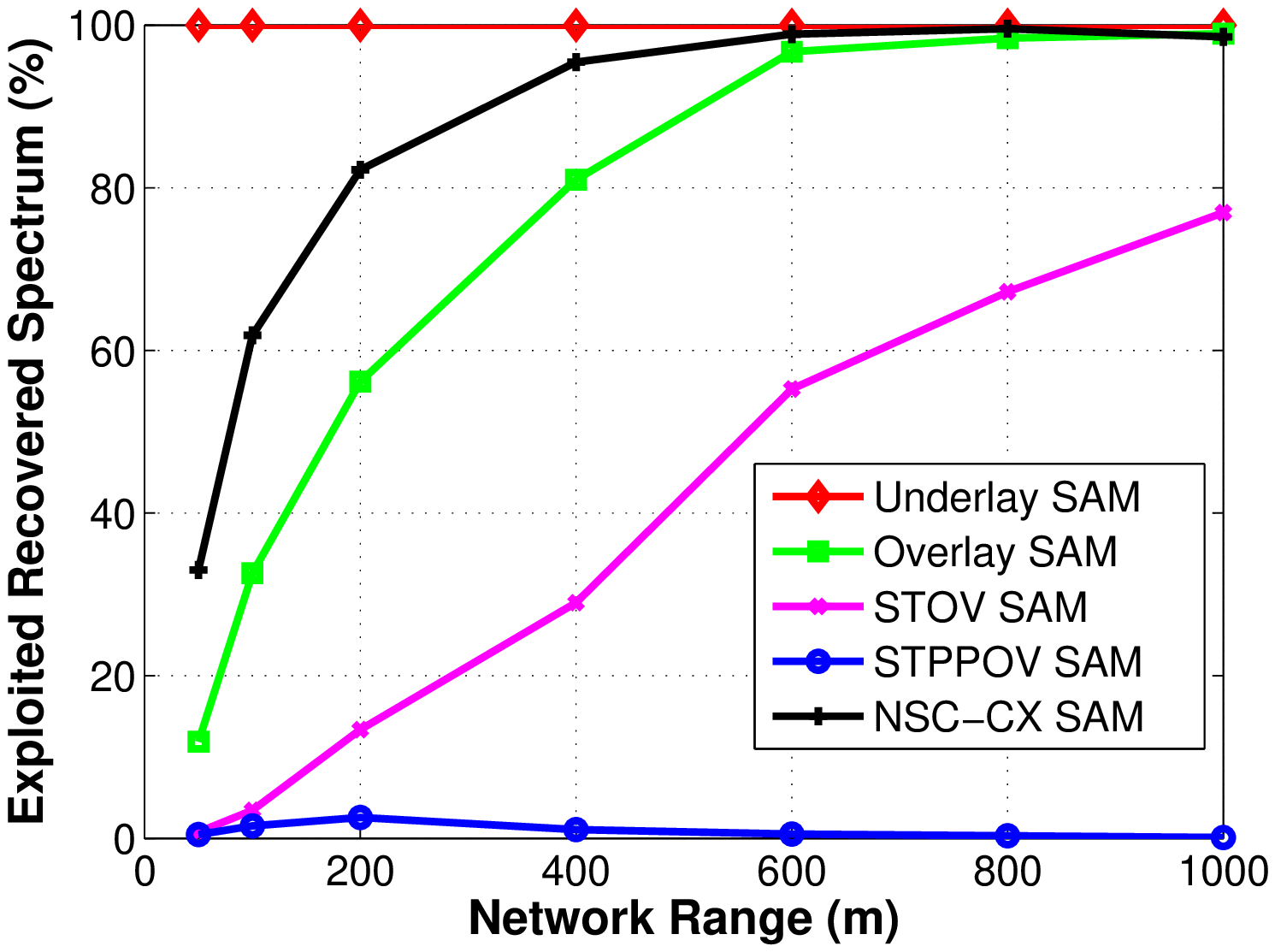}}
{\includegraphics [width=0.454\textwidth, angle=0] {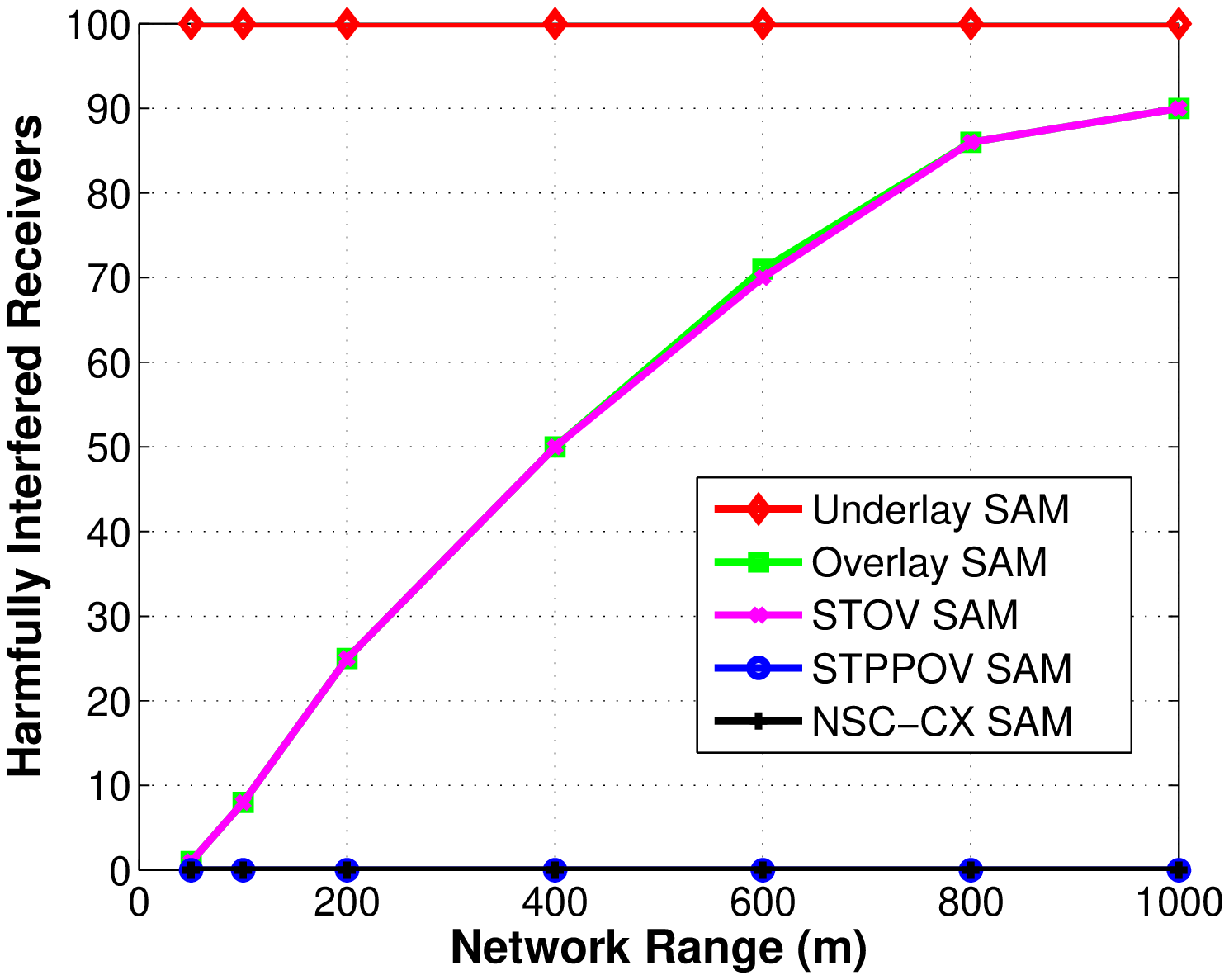}}
{\includegraphics [width=0.454\textwidth, angle=0] {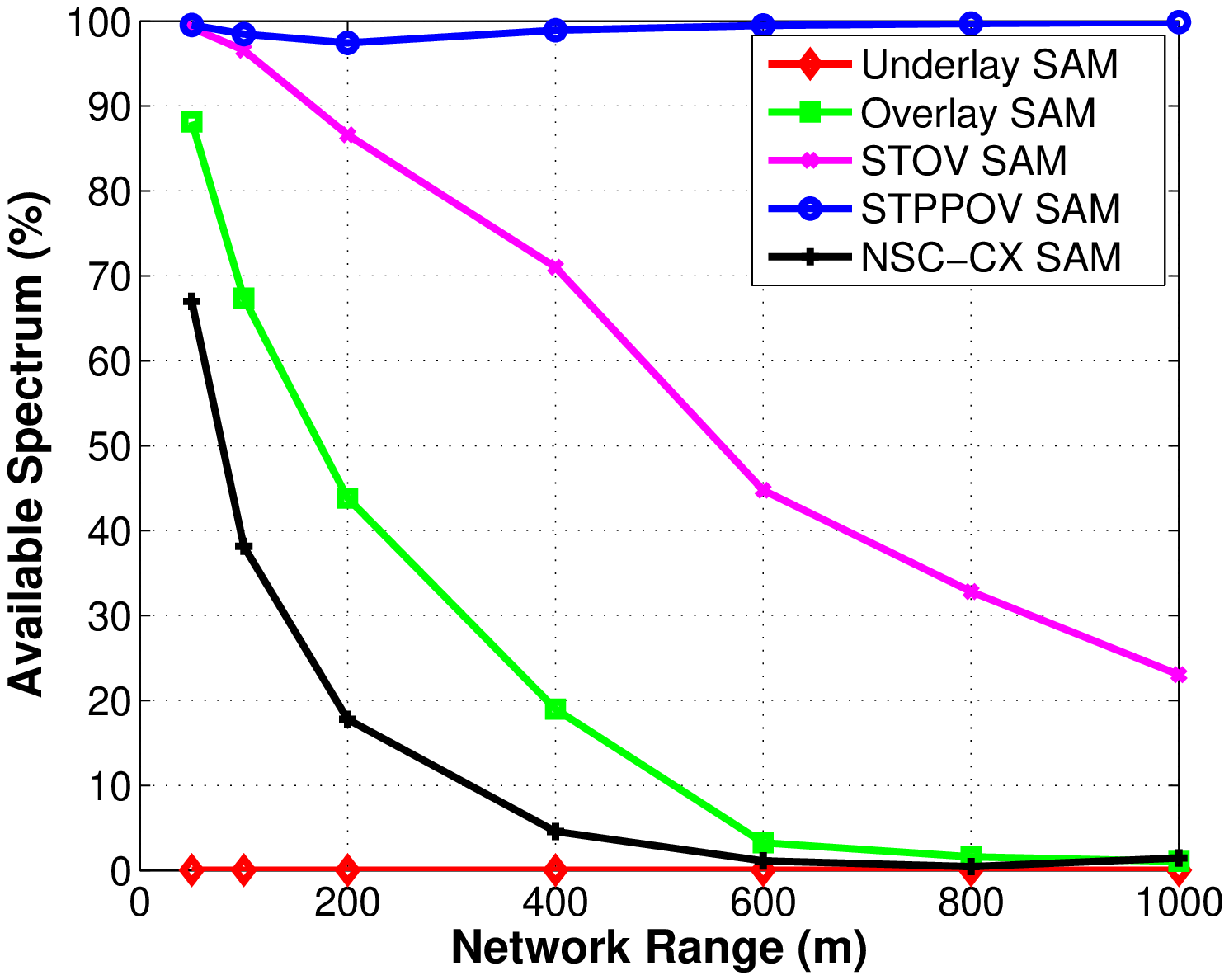}}
{\includegraphics [width=0.454\textwidth, angle=0] {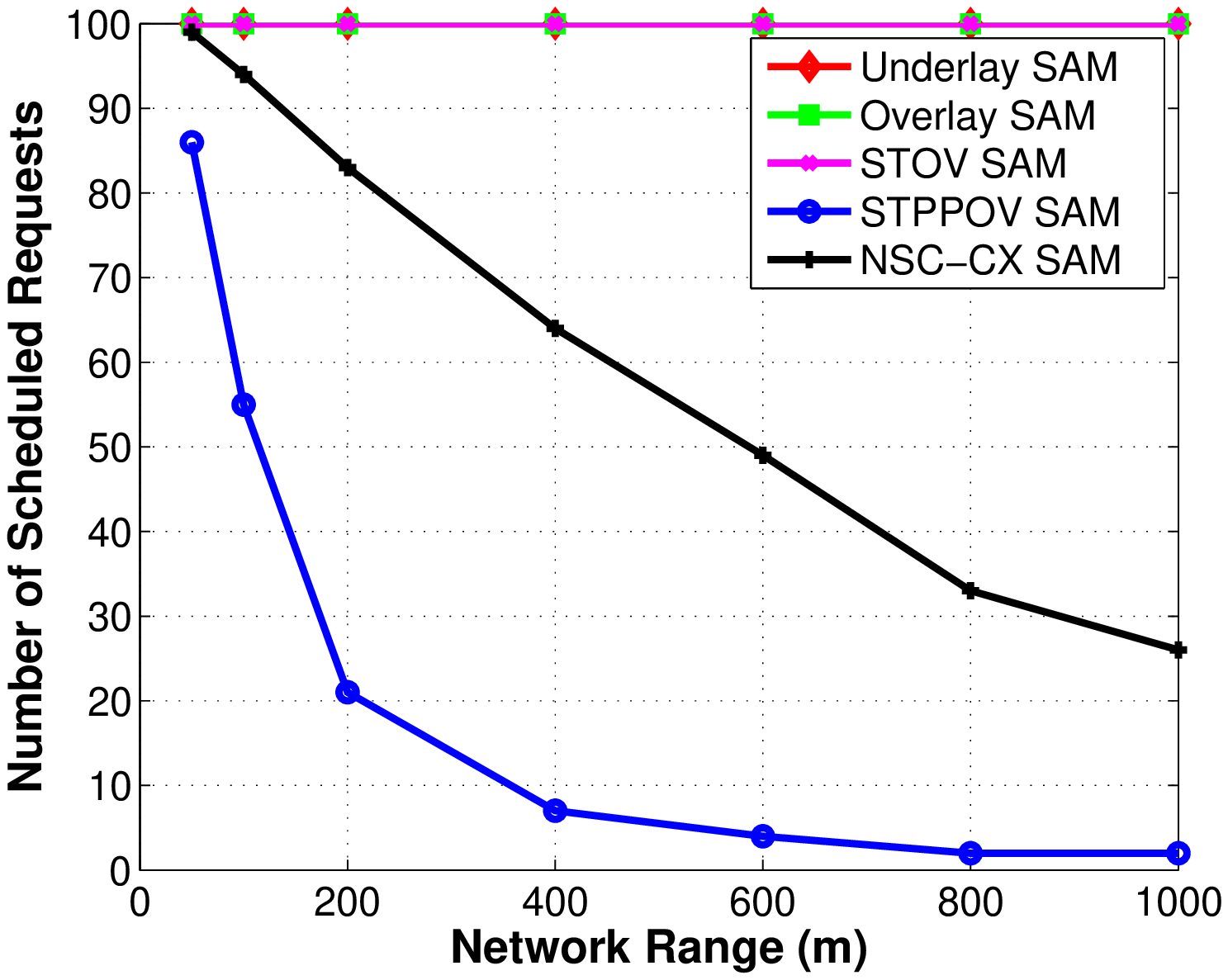}}
\caption{The performance comparison of SAMs with the varying network range \textbf{(Experiment-6)}. In this experiment, we observe that fine granular access with smaller network-ranges leads to a higher number of successful spectrum-access requests. }
\label{fig:L502_ST9}
\end{figure}

In this experiment, the performance of various SAMs is characterized with the varying range of the networks in the dynamic spectrum sharing scenario. All the networks are randomly positioned. The minimum desired SINR at the receivers is 3 dB. The transmitters and receivers are directional with $60^{\circ}$ antenna beamwidth. From Figure~\ref{fig:L502_ST9}, we observe that
\begin{itemize}
  \item The smaller network ranges lead to a larger number of spectrum-access requests being satisfied. The SINR and the hence the throughput is also high in this case. \textit{We argue that the fine granular spectrum access needs to be adopted whenever possible.}
	\item This scenario helps us to understand the expected performance in terms of the number of successfully scheduled spectrum-access requests when larger network ranges are desired. The performance of `NSC-CX SAM' that exploits the knowledge of all transceiver positions and prioritizes the spectrum-access requests by spectrum-consumption costs is reasonably good for all network ranges; Especially, for the larger network ranges, `NSC-CX SAM' schedules a larger number of the spectrum-access requests as compared to other SAMs. 
\end{itemize}

\section{Conclusions}
In this paper, we investigated the potential of spectrum sharing in order to make it non-harmful, efficient, and therefore feasible from the perspective of incumbents. In order to improve the potential of secondary spectrum access, we studied how the availability and the exploitation of the RF spectrum could be maximized. Our findings suggest the need for 
\begin{itemize}
  \item \textbf{Fine granular spectrum-access policy:} Spectrum is consumed by transmitters as well as receivers. Thus, spectrum-accesses within a RF-system may vary significantly in terms of their demand for RF-spectrum. \textit{Defining a spectrum-access policy at the granularity of individual spectrum-access request} enables us to share and exploit the available spectrum space more efficiently.
  \item \textbf{Active role of incumbents:} By altering the primary spectrum-access attributes and incorporating the knowledge of spectrum-access attributes of the primary transceivers, the spectrum available for secondary access can be significantly improved.
  \item \textbf{Exploiting fine granular spatial reuse opportunities:}. A large portion of the spectrum-access opportunities are fine granular in the space, time, and frequency dimensions. Thus, the ranges of the spectrum-access requests play a crucial role in the SAM's ability to exploit these fine granular spectrum-access opportunities. The throughput and the number of scheduled spectrum-access requests are significantly improved with a fine granular approach to spectrum access.  
	\item \textbf{Quantified spectrum-access footprints:}  The spectrum consumed by the transmitters and the receivers can be precisely controlled with allocation of the \textit{quantified} spectrum-access footprints. This helps to improve the number of scheduled spectrum-access requests.
  \item \textbf{Network-coexistence with non-harmful interference.}  It is necessary to ensure minimum throughput for secondary spectrum access requests in order to make it attractive for designing services based on secondary access to spectrum. Thus, it is necessary for a SAM to ensure non-harmful interference among \textbf{\textit{all}} spectrum sharing networks.
	 \item \textbf{Adopting interference-tolerant signal model and transceiver technology:} A higher minimum SINR for successful reception of the signal under cochannel interference conditions implies a higher consumption of spectrum by the receivers.  Adopting interference-tolerant modulation format and transceiver technology with higher interference-tolerance can help to improve the available spectrum. Also, directional transmission and reception abilities are desired in order to further reduce the spectrum consumption by transceivers.
\end{itemize}

The top-level conclusion of this study is secondary access to spectrum could be efficient while protecting the existing spectrum uses. \textit{We demonstrated that close to 100 spatially overlapping heterogeneous spectrum-access requests can be scheduled in a small 4.3 $km$ x 3.7 $km$ geographical region} using the suboptimal approach.  By playing an active role in secondary spectrum access, an incumbent can extract significant value out of the underutilized spectrum while protecting their primary service networks. The potential business opportunities could be able to justify the cost of the infrastructure necessary for spectrum consumption management.

The uncertainties in the radio environment hinder the potential of a SAM to efficiently share the \textit{available} spectrum. We observed a significant amount of the available spectrum is \textit{lost} due to conservative assumptions regarding the propagation environment. In this regard, there is a need for fine granular characterization of the propagation environment that would enable more accurate estimation of the available spectrum in real-time. This would also help in the enforcement of quantified spectrum-access policy and facilitate real-time dynamic spectrum sharing.




\end{document}